\documentclass[reprint,aps,pre]{revtex4-2}
\usepackage[utf8]{inputenc}
\usepackage[T1]{fontenc}
\usepackage{amsmath}
\usepackage{amssymb}
\usepackage{graphicx}
\usepackage{natbib}
\usepackage[hidelinks]{hyperref} \usepackage{cleveref}
\usepackage[caption=false]{subfig}
\usepackage{mathtools}

\makeatletter
\newcommand{\getfont}{\f@family~\f@size pt}
\makeatother

\newcommand{\erf}{\operatorname{erf}}

\newcommand{\dd}{\mathrm{d}}
\renewcommand{\vec}{\mathbf}
\newcommand{\pder}[2]{\frac{\partial #1}{\partial #2}}
\newcommand{\curl}{\nabla\times}
\renewcommand{\div}{\nabla\cdot}
\renewcommand{\Re}[1]{\mathrm{Re}\left\{#1\right\}}
\renewcommand{\Im}[1]{\mathrm{Im}\left\{#1\right\}}

\makeatletter\begin{document}

\title{Current-driven Langmuir Oscillations and Streaming Instabilities}
\author{Sigvald Marholm}
\email{sigvald@marebakken.com}
\homepage{https://sigvaldm.github.io}
\affiliation{Department of Physics, University of Oslo, P.O. Box 1048 Blindern, N-0316 Oslo, Norway}
\affiliation{Present affiliation: Department of Computational Materials Processing, Institute for Energy Technology, Instituttveien 18, N-2007 Kjeller, Norway}
\author{Sayan Adhikari}
\affiliation{Department of Physics, University of Oslo, P.O. Box 1048 Blindern, N-0316 Oslo, Norway}
\author{Wojciech J. Miloch}
\affiliation{Department of Physics, University of Oslo, P.O. Box 1048 Blindern, N-0316 Oslo, Norway}
\date{\today}

\begin{abstract}
    The Buneman and ion acoustic instabilities are usually associated with different electron and ion drift velocities, in such a way that there is a large current through the plasma.
    However, due to the recently discovered \emph{current-driven Langmuir oscillations} \cite{baumgaertel,sauer2015,sauer2016}, the relative drift velocity in these configurations will oscillate at the plasma frequency, and with an amplitude of at least the initial drift velocity.
    In contrast, the textbooks assume a \emph{constant} drift velocity. Since the growth rates arrived at under that assumption are far less than the plasma frequency, several oscillation periods will take place during the linear growth phase, and this will dampen the instabilities. We provide general theoretical derivations of these oscillations, and show simulation results of the altered behavior of the instabilities. Towards the end, we hypothesize that \emph{drift-averaging} might be a viable method of calculating the modified growth rates.
\end{abstract}
 
\maketitle

\section{Introduction}

\begin{figure}
     \centering
     \subfloat[The two-stream instability (current-free)]{
         \includegraphics{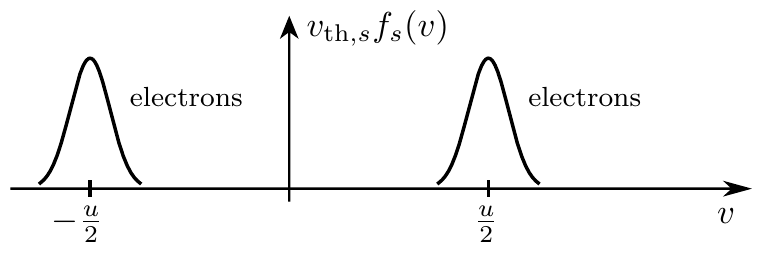}
         \label{fig:config_symmetric}
     }
     
     \subfloat[The two-stream and Buneman instabilities]{
         \includegraphics{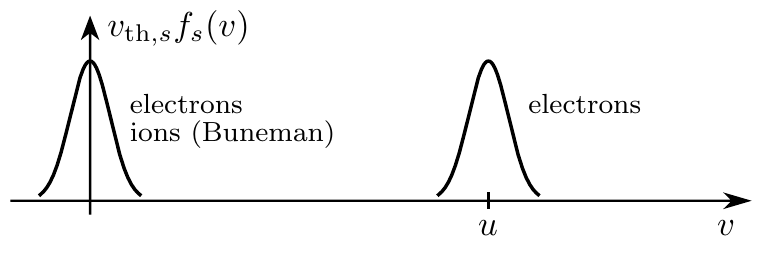}
         \label{fig:config_asymmetric}
     }

     \subfloat[The ion acoustic instability]{
         \includegraphics{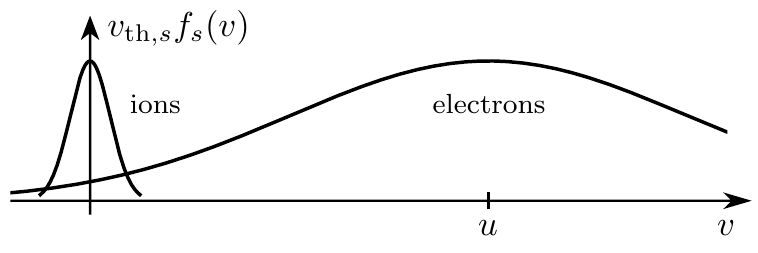}
         \label{fig:config_ionac}
     }
    \caption{Initial velocity distributions for the instabilities under consideration. $f_s(v)$ is the distribution of velocities $v$ of the species $s$. The distributions are here scaled to equal peak values by multiplying with the thermal speed $v_{\mathrm{th},s}$.}
    \label{fig:config}
\end{figure}

Some of the most studied plasma instabilities are the so-called
\emph{two-stream instabilities}, where there exist a difference in drift
velocity between two cold species \cite{treumann_advanced,hasegawa,chen}. These
can either be the same species, like in the \emph{electron--electron two-stream
instability}, or they can be different, like in the \emph{ion--electron
two-stream instability}, also known as the \emph{Buneman instability}. Possible
initial scenarios are illustrated in a 1D velocity space in
\cref{fig:config_symmetric,fig:config_asymmetric}. According to the textbooks,
small perturbations will grow exponentially in these systems, with a growth
rate ranging from about $0.05\omega_{pe}$ to $0.5\omega_{pe}$. Another common
textbook example is the \emph{ion acoustic instability}, where the electron
temperature is much larger than the ion temperature (\cref{fig:config_ionac}).
The predicted growth rate is approximately $0.05\omega_{pi}$ in this case. For
the Buneman and ion acoustic instabilities, there is a strong current which
causes some effects which are usually not accounted for in the literature. For
the electron--electron two-stream instability, it is possible to sidestep this
problem by choosing a current-free, symmetric reference frame like the one in
\cref{fig:config_symmetric}. However, this is not possible for the Buneman and
ion acoustic instabilities.

One effect caused by the current is the induction of an azimuthal magnetic
field around the streaming electrons. This would cause a pinch effect
\citep{hasegawa,pecseli}, and likely also affect the development of the
instabilities. However, when studying the Buneman and ion acoustic
instabilities in the electrostatic regime (as is customary), this effect is
ignored. This can be justified for instance when considering electrons
streaming along magnetic field lines of the Earth, or in a plasma device, such
that the azimuthal component becomes negligible compared with the longitudinal
component of the magnetic field. Henceforth, we too shall ignore this effect.

Another effect, is that even in the electrostatic approximation, the current
makes the drift velocity $u$ oscillate back and forth across zero at the plasma
frequency, and with an amplitude of at least the initial velocity $u_0$.
Curiously, such \emph{current-driven Langmuir oscillations} have only recently
been discussed, first in 2013 by \citet{baumgaertel}, and then in 2015--2016 by
\citet{sauer2015,sauer2016}, who showed that these oscillations modulate other
waves. To our knowledge, we are the first to discuss the effect these
oscillations have on the streaming instabilities. Since the Langmuir
oscillation frequency is much larger than the growth rates found when assuming
a constant drift velocity, the traditionally derived growth rates do not hold.
In fact, we will show that several periods take place within the linear growth
phase, and that the instability grows at reduced rate, if at all, because of that.

We begin in \cref{sec:background} with a review of the two-stream and ion
acoustic instabilities, as usually derived in textbooks. In
\cref{sec:langmuir}, we use Ampére's equation to show that the strong current
in these systems lead to a form of Langmuir oscillations, similar to what is
presented by \citet{baumgaertel,sauer2015,sauer2016}, although we generalize it
to arbitrary velocity distributions. While these papers emphasized the
importance of using Ampére's law instead of Poisson's equation, we go further
in \cref{sec:helmholtz}, and explain \emph{why} using Poisson's equation fails
for these electrostatic oscillations. In \cref{sec:simulations}, we show
numerical evidence for the oscillations, both occurring alone, and co-existing
with two-stream or ion acoustic instabilities. The simulations are carried out
by solving the Vlasov-Maxwell equations using the Gkeyll code \cite{gkeyll}. In
\cref{sec:modified}, we present a method of averaging the growth rates over
changing drift velocities, and get good agreement with growth rates obtained
from simulations. The conclusion follows in \cref{sec:conclusion}.

 \section{Background}
\label{sec:background}

\subsection{Electrostatic instabilities}
\label{sec:beamplasma}
Let us revisit the linear theory for 1D electrostatic plasma instabilities. First, the Vlasov-Poisson equations are linearized, and then all field and velocity perturbations are assumed to be waves, proportional to
$e^{i(kx-\omega t)}$, such that the linearized Vlasov-Poisson equations turn from differential to algebraic form. Eliminating the field and velocity amplitudes from these equations leads to a relation between $\omega$ and $k$ that must be satisfied by any solution. Assuming the unperturbed electric field to be zero, this \emph{dispersion relation} becomes
\begin{align}
    \varepsilon(\omega, k) = 1 + \sum_s\chi_s(\omega, k) = 0.
    \label{eq:generic_disprel}
\end{align}
$\varepsilon$ is actually the relative permittivity due to the plasma, which can be seen by comparing the linearized Poisson's equation in microscopic form with the macroscopic form. Moreover, the response due to vacuum is the ``1'' in the above equation, while $\chi_s$ is the response (or electric susceptibility) due to species $s$,
\begin{align}
    \chi_s(\omega, k) = \frac{q_s^2}{\varepsilon_0 m_s}
    \int\limits_{-\infty}^\infty \frac{\partial f_{s0}/\partial v}{(\omega/k-v)} \,\dd v.
\label{eq:generic_response}
\end{align}
Here, $f_{s0}(x,v)$ is the initial, unperturbed distribution of the species in phase space $(x,v)$, and $q_s$ and $m_s$ is its charge and mass, respectively. $\varepsilon_0$ is the vacuum permittivity.
For a Maxwellian species, the response can be written
\begin{align}
\chi_s(\omega, k) = -\frac{1}{2k^2\lambda_{Ds}^2} Z'\bigg(\underbrace{\frac{\omega/k-u_s}{\sqrt{2}v_{\mathrm{th},s}}}_{\zeta_s}\bigg),
    \label{eq:maxwellian_response}
\end{align}
where $Z$ is the plasma dispersion function, and $Z'$ is its derivative \cite{fitzpatrick,cagas}.
Solutions to \cref{eq:generic_disprel} (a.k.a. modes) where $\omega$ is complex, $\omega=\omega_r+i\gamma$, have unstable exponential growth when $\gamma>0$, or dampens out when $\gamma<0$. It is also possible to have complex-valued $k$ -- so-called \emph{convective instabilities} -- but in this paper we focus on modes with real $k$, i.e., \emph{absolutely unstable} ones.

Before we go on, it is instructive to identify which dimensionless groups or factors are at play (c.f. Buckingham's $\pi$-theorem \cite{buckingham}). For each species, there are only three independent dimensionless groups to be found in \cref{eq:maxwellian_response}, for instance
\begin{align}
    \frac{\omega}{\omega_{ps}} ,\quad
    k\lambda_{Ds} ,\quad
    \frac{u_s}{v_{\mathrm{th},s}}.
\end{align}
Thus for two species $s\in\{1,2\}$ (which cover all the cases in this article), the dispersion relation can be written solely in terms of six dimensionless groups. Rather than just repeating the above three groups twice, it is convenient to refactor them, and if we also consider the two species to have equal density and charge (up to sign), we can refactor them to
\begin{align}
    \frac{\omega}{\omega_{p1}} ,\quad
\frac{ku_1}{\omega_{p1}} ,\quad
    \frac{m_2}{m_1} ,\quad
    \frac{v_{\mathrm{th},2}}{v_{\mathrm{th},1}} ,\quad
    \frac{u_1}{v_{\mathrm{th},1}} ,\quad
    \frac{u_2}{v_{\mathrm{th},1}}.
    \label{eq:groups}
\end{align}
Notice that the first two groups represent a normalized frequency and wavenumber. 
Another natural choice for the wavenumber would have been $k\lambda_{D1}$. However, this becomes nonsensical in the cold limit (where $\lambda_{D1}=0$) which is important for the instabilities we study. Interestingly, the sign of the charges does not enter any of the groups.

The implication of these groups is that the normalized frequency, including its imaginary part the normalized growth rate $\gamma/\omega_{p1}$, do not depend on parameters such as density and drift velocity separately, but only on the other dimensionless groups. By reporting growth rates (and other values) in terms of these groups we thus cover entire classes of situations with different densities, drift velocities, etc. The cases we study in this paper, with respect to the latter four groups, are given in \cref{tab:parameters}.

\begin{table*}
    \centering
    \caption{Cases studied in this paper}
    \begin{tabular}{lrrrr}
        \hline
& $\frac{m_2}{m_1}$ & $\frac{v_{\mathrm{th},2}}{v_{\mathrm{th},1}}$ &
            $\frac{u_1}{v_{\mathrm{th},1}}$ & $\frac{u_2}{v_{\mathrm{th},1}}$   \\
        \hline
        Current-free two-stream (electron--electron) & 1   & 1      & 5 & $-5$  \\
        Two-stream (electron--electron)              & 1   & 1      & 10 & 0    \\
        Buneman (ion--electron)                      & 100 & 1      & 10 & 0    \\
Ion acoustic (ion--electron)                 & 100 & 0.0125 & 0.25 & 0 \\
        \hline
    \end{tabular}
    \label{tab:parameters}
\end{table*}

Next, let us provide some limiting forms of \cref{eq:maxwellian_response}. For species with large arguments, $|\zeta_s|\gg 1$, the response may be simplified by series expansion of $Z$ \cite{fitzpatrick}:
\begin{align}
    \chi_s(\omega, k) &\approx -\frac{\omega_{ps}^2}{(\omega-ku_s)^2} + i \frac{1}{k^2\lambda_{Ds}^2}\sqrt{\pi}\zeta_s e^{-\zeta_s^2}
    \label{eq:large_argument}
\end{align}
The most important application of this approximation is to cold species, i.e., when $v_{\mathrm{th},s}\rightarrow 0$. As a matter of fact, if \cref{eq:generic_disprel} was derived using the momentum equation for a cold fluid (i.e. with zero pressure) instead of the Vlasov equation, $\chi_s$ would coincide with the above expression, except it would be missing the imaginary part (which in any case is small because $|\zeta_s|\gg 1$).

For species with small arguments, $|\zeta_s|\ll 1$, a series expansion yields the following simplified response:
\begin{align}
    \chi_s(\omega, k) &\approx \frac{1}{k^2\lambda_{Ds}^2} + i \frac{1}{k^2\lambda_{Ds}^2}\sqrt{\pi}\zeta_s e^{-\zeta_s^2}
    \label{eq:small_argument}
\end{align}
For many cases, e.g., when deriving the growth rates for the two-stream and Buneman instabilities, the imaginary parts of the large-argument and small-argument approximations can be omitted. This then leads to a purely real dispersion relation $\varepsilon(\omega, k)=0$, but which still has complex roots and hence a non-zero growth rate. In other cases, such as for the ion acoustic instability, omitting the imaginary parts of $\chi_s$ leads to a dispersion relation with real roots only. The ion acoustic instability can thus be captured only by including the imaginary parts of $\chi_s$.

 \subsection{Two-stream instabilities}
\label{sec:buneman}
Two-stream instabilities occur when one species (subscript 1) is streaming with respect to another species (subscript 2) with a velocity $u$, and their thermal speeds are much less than $u$. In this case their distributions do not overlap, which permits the use of the cold fluid approximation, i.e., \cref{eq:large_argument}. Taking species 2 to have zero drift velocity, like in \cref{fig:config_asymmetric}, the dispersion relation can be written
\begin{align}
\varepsilon(\omega, k) =
    1 - \frac{\omega_{p2}^2}{\omega^2} - \frac{\omega_{p1}^2}{(\omega- ku)^2} = 0.
    \label{eq:disprel_asymmetric}
\end{align}
For the electron--electron two-stream instability, the two species are both electrons, and $\omega_{p1}=\omega_{p2}=\omega_{pe}$, whereas for the ion--electron two-stream instability or Buneman instability, species 2 is an ion species ($\omega_{p1}=\omega_{pe}$ and $\omega_{p2}=\omega_{pi}$).
Interestingly, changing the charge polarity of a species does not change the dispersion relation. The only difference between the dispersion relation for the electron--electron two-stream instability (hereinafter just referred to as ``the two-stream case'') and the Buneman instability is thus the mass ratio between the species.

\begin{figure*}
    \centering
    \subfloat[The two-stream case]{
        \includegraphics{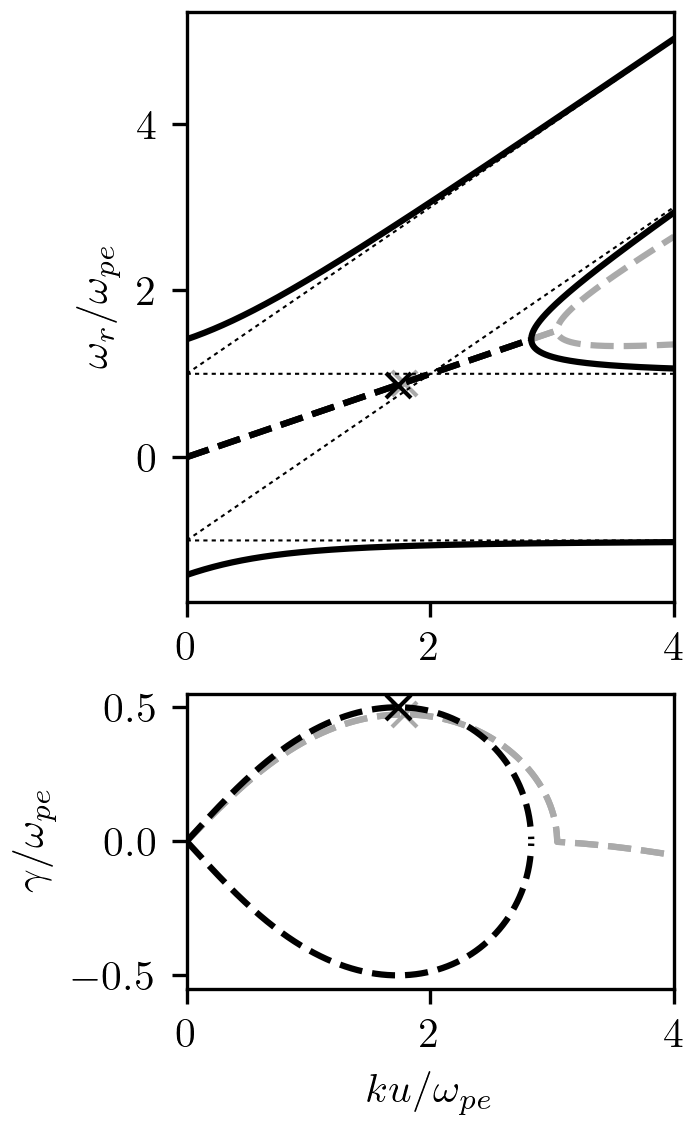}
        \label{fig:disprel_twostream}
    }
    \subfloat[The Buneman case]{
        \includegraphics{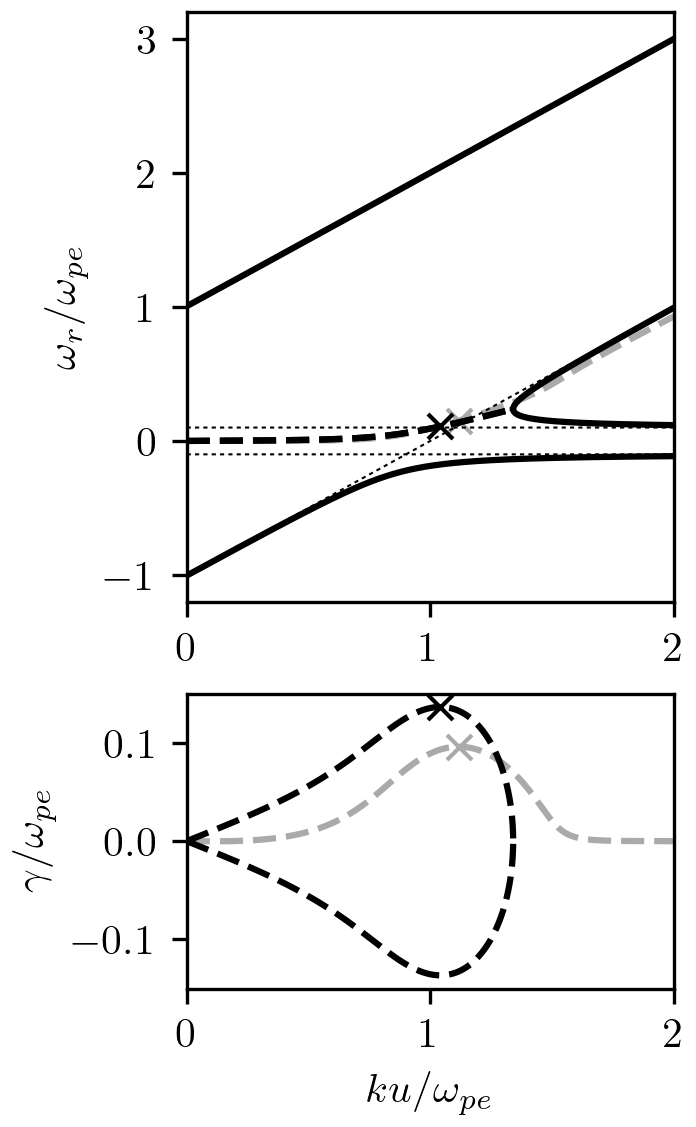}
        \label{fig:disprel_buneman}
    }
    \subfloat[The ion acoustic case]{
        \includegraphics{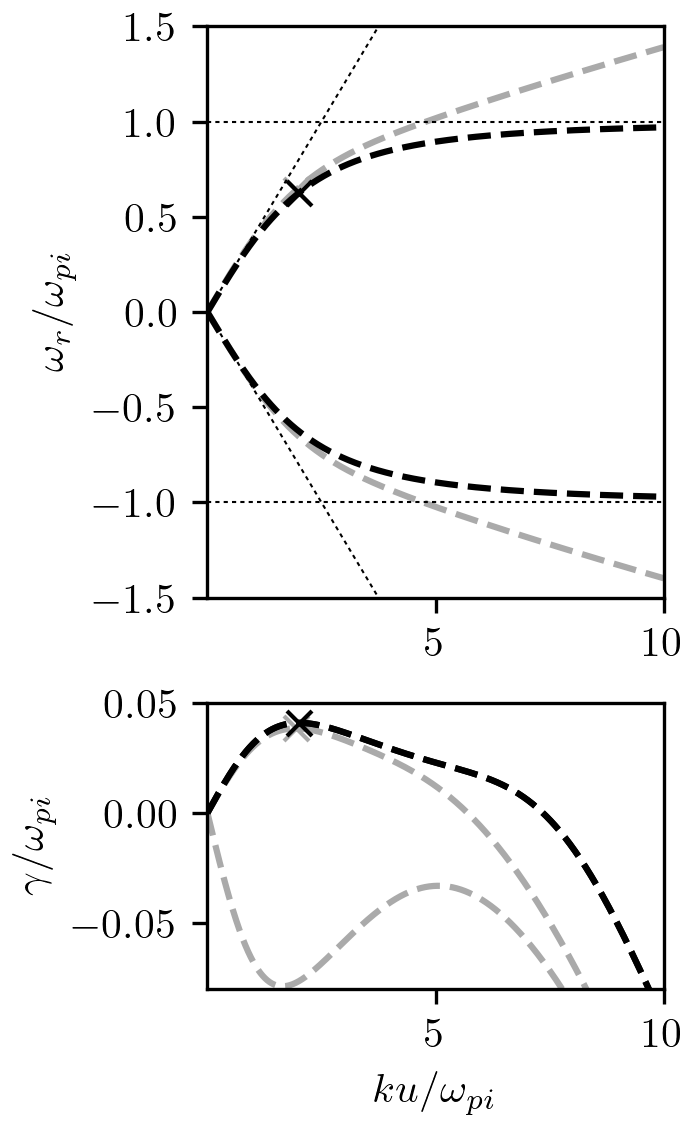}
        \label{fig:disprel_ionac}
    }

    \caption{
        The solid black lines in the upper plots show the real roots of $\varepsilon(\omega, k)$, according to approximate analytic expressions. When the roots become complex, they are plotted as \emph{dashed} black lines instead, with the real part in the upper plot, and the imaginary part in the lower plot. The thin, dotted lines in the upper plots are asymptotes of the analytic expressions. The gray lines are numerically obtained roots of the more accurate kinetic dispersion relation, \cref{eq:disprel_asymmetric_kinetic}. Finally, the crosses indicate the fastest-growing modes, which are obtained by picking the root which have the largest imaginary part. The values of these modes are given in \cref{tab:twostream,tab:buneman,tab:ionac}}
\label{fig:disprel}
\end{figure*}

For a given $k$, the roots $\omega$ of $\varepsilon(\omega, k)$ can be obtained by multiplying it by all its denominators, and using a standard polynomial root finder (e.g. \texttt{numpy.roots} in NumPy \cite{numpy}). This is done for a sweep of wavenumbers for both the two-stream case and the Buneman case in \cref{fig:disprel_twostream,fig:disprel_buneman} (black), respectively. For short wavelengths (large $k$), the four roots are real, and approach the asymptotic trends obtained by considering the dynamics of the species independently. In particular, the asymptotes for species 2 can be obtained by discarding the term for species 1 from \cref{eq:disprel_asymmetric}, resulting in $\omega=\pm\omega_2$ (negative $\omega$ indicate that the wave propagates in the negative direction). Discarding instead the term for species 2, we obtain the sloped asymptotes $\omega=ku\pm\omega_1$ for species 1. The upper sloped asymptote represent the so-called \emph{fast waves} (due to their higher phase velocities $\omega/k$), while the lower represent the \emph{slow waves}. Unlike waves in the other branches, the slow waves carry negative energy (as seen from our reference frame) \cite{hasegawa}. As the wavelength increases and $k$ is lowered, the dynamics of the two species couple, which can be seen from the merging of the slow, negative-energy wave with the positive-energy wave of the stationary species in \cref{fig:disprel_twostream,fig:disprel_buneman}. Below the merging point $ku/\omega_{pe}=2.83$ (two-stream) or 1.34 (Buneman, values obtained from plotting script), two real roots transitions into a complex-conjugate pair.

It is interesting to observe that for the two-stream case, it is possible to move to a current-free frame of reference, by placing the two electron distributions symetrically about the origin in velocity space (c.f. \cref{fig:config_symmetric}).
This change of reference frame lead to a Doppler shift, which can be taken into account by the substitution $\omega\rightarrow\omega+0.5ku$ in \cref{eq:disprel_asymmetric}.
The effect is that $0.5ku$ gets subtracted from the real part of the roots depicted in \cref{fig:disprel_twostream}, thereby ``tilting'' them downwards so they become symmetric about the $x$-axis, while the imaginary part remains unchanged.
The Buneman case cannot be made current-free by a change of reference frame.

It is the wave mode with the largest imaginary part $\gamma$ that exhibits the fastest unstable growth $e^{\gamma t}$, and that will first arise from random perturbations. This mode will grow in amplitude, overshadowing the other modes, until the linearizations in the dispersion relation becomes invalid. At that point we enter the non-linear regime, where the plasma thermalizes and turbulence may occur \cite{treumann_advanced}. The wavenumber, frequency and growth rate of the fastest-growing mode can be calculated analytically for the two-stream case,
\begin{align}
    \frac{ku}{\omega_{pe}}=\sqrt{3}, \quad
    \frac{\omega}{\omega_{pe}} = 0.5(\sqrt{3}+i),
    \label{eq:fastest_growing_twostream}
\end{align}
whereas a common approximation for the Buneman instability is:
\begin{align}
    \frac{ku}{\omega_{pe}} \approx 1, \quad
    \frac{\omega}{\omega_{pe}} \approx \left(\frac{1}{16}\frac{m_e}{m_i}\right)^\frac{1}{3}(1+i\sqrt{3}).
\label{eq:fastest_growing_buneman}
\end{align}
(See \cref{app:growthrate} for derivations.) However, for the Buneman case, more accurate values can be obtained simply by identifying the fastest growing mode in the plotting script. For convenience, these numbers are listed in \cref{tab:twostream} for the two-stream instability, and \cref{tab:buneman} for the Buneman instability, and compared with numbers we are yet to discuss.

\begin{table}
    \caption{Characteristics of the fastest-growing wave for the two-stream instability}
    \begin{tabular}{lrrr}
        \hline
        & $\frac{ku}{\omega_{pe}}$ & $\frac{\omega_r}{\omega_{pe}}$ & $\frac{\gamma}{\omega_{pe}}$ \\
        \hline
        Fluid theory (\cref{eq:fastest_growing_twostream} or \cref{fig:disprel_twostream})                             & 1.73 & 0.866 & 0.500 \\
Kinetic theory                            & 1.78 & 0.892 & 0.473 \\
        \hline
        Simulation (current-free)                 & 1.78 &       & 0.472 \\
        Simulation (with current)                 & 1.78 &       & 0.461 \\
        \hline
    \end{tabular}
    \label{tab:twostream}
\end{table}

\begin{table}
    \caption{Characteristics of the fastest-growing wave for the Buneman instability}
    \begin{tabular}{lrrr}
        \hline
        & $\frac{ku}{\omega_{pe}}$ & $\frac{\omega_r}{\omega_{pe}}$ & $\frac{\gamma}{\omega_{pe}}$ \\
        \hline
        Fluid theory (\cref{eq:fastest_growing_buneman}) & 1.00 & 0.085 & 0.148 \\
        Fluid theory (\cref{fig:disprel_buneman})   & 1.04 & 0.108 & 0.136 \\
Kinetic theory                                   & 1.12 & 0.153 & 0.096 \\
        \hline
        Simulation                                       & 1.12 &       & 0.036 \\
        \hline
        Drift-averaged method                            &      &       & $-$0.021 \\
        Drift-averaged method (mode hopping)             &      &       & 0.038 \\
        \hline
    \end{tabular}
    \label{tab:buneman}
\end{table}

Unfortunately, it is impractical to make the thermal speeds very small in our simulations, owing to finite velocity space resolutions. In the simulations, we let $v_{\mathrm{th},s}/u=0.1$ (for both species), which will lead to some degree of Landau damping and thus lower growth rates than predicted using the cold fluid approximation.
To account for the Landau damping also in theory, we must consider the full, kinetic dispersion relation:
\begin{align}
    \varepsilon(\omega, k) = 1
    &-\frac{1}{k^2\lambda_{D2}^2}Z'\left(\frac{\omega/k}{\sqrt{2}v_{\mathrm{th},2}}\right) \nonumber\\
    &-\frac{1}{k^2\lambda_{D1}^2}Z'\left(\frac{\omega/k-u}{\sqrt{2}v_{\mathrm{th},1}}\right)=0.
    \label{eq:disprel_asymmetric_kinetic}
\end{align}
This dispersion relation has infinitely many solutions $\omega$ for each $k$, and they are not as straight-forward to obtain as for \cref{eq:disprel_asymmetric}.
However, with a good initial guess, it is possible to use a numerical root finder.
We simply use the secant method (\texttt{scipy.optimize.newton} in SciPy \cite{scipy}), with the previously obtained solutions of \cref{eq:disprel_asymmetric} as initial guesses.
Selected kinetic roots of interest that correspond to the fluid ones are depicted in gray in \cref{fig:disprel_twostream,fig:disprel_buneman}, and characteristics of the fastest-growing mode are listed in \cref{tab:twostream,tab:buneman}.
 \subsection{The ion acoustic instability}
\label{sec:ionacoustic}
A uniform, Maxwellian ion--electron plasma supports the propagation of so-called \emph{ion acoustic waves} \cite{fitzpatrick}. However, if the ions and electrons have no relative drift, these waves are Landau dampened. In fact, unless the electron temperature is much larger than the ion temperature ($T_e\gg T_i$), the waves are so strongly dampened that they will not even propagate a few wavelengths before being indiscernible, which is why ion acoustic waves are usually observed at large temperature ratios \cite{fitzpatrick}.

\citet{treumann_advanced} considers two modifications that turn these waves into \emph{ion acoustic instabilities}, with positive growth rates. The first is to displace the electron distribution in velocity space, such that it has a relative drift $u$ with respect to the ions (\cref{fig:config_ionac}). The second is to introduce a third species: a low-density ion beam passing through the stationary ion--electron plasma. The latter configuration is essentially current-free, and thus not affected by the current-driven Langmuir oscillations central to this work. We therefore limit our discussion to the first kind, which also bears more resemblance to the Buneman instability as depicted in \cref{fig:config_asymmetric}.

The full dispersion relation is again given by \cref{eq:disprel_asymmetric_kinetic}, although with a larger thermal speed for the electrons than in the previous cases. It is commonplace, however, to consider a simplified, approximate form \cite{treumann_advanced}.
The Penrose criterion dictates that a necessary criterion for (weak) instabilities is \cite{hasegawa}
\begin{align}
    \left. \frac{\omega}{k}\frac{\partial f_s}{\partial v} \right|_{v=\omega/k}>0.
\end{align}
The only phase velocities that satisfy this (with respect to \cref{fig:config_ionac}) are those between the ion and electron peaks. In this region, $\partial f_i/\partial v$ is still negative, which means that the ions still Landau dampens the wave. However, $\partial f_e/\partial v>0$, meaning that the electrons feed their free energy into the wave. If the growth due to the electrons is larger than the damping due to the ions, we have an inverse Landau damping-type growth. To lessen the ion damping, we seek waves of phase velocity $\omega/k\gg v_{\mathrm{th},i}$, such that the slope has flattened out. This permits us to use the large-argument approximation, \cref{eq:large_argument}, for the ions. For the electrons, we assume $\omega/k$ to not be too far from the electron peak, i.e., $u-\omega/k\ll v_{\mathrm{th,e}}$, which allow us to use the small-argument approximation, \cref{eq:small_argument}. The dispersion relation then takes the form
\begin{align}
    \varepsilon(\omega, k) = \varepsilon_r(\omega, k) + i\varepsilon_i(\omega, k)=0,
\end{align}
where
\begin{align}
    \varepsilon_r(\omega, k) = 1 + \frac{1}{k^2\lambda_{De}^2} - \frac{\omega_{pi}^2}{\omega^2},
\end{align}
and
\begin{align}
    \varepsilon_i(\omega, k) = \frac{\sqrt{\pi}}{k^2\lambda_{De}^2}
    \left( \zeta_e e^{-\zeta_e^2} + \frac{T_e}{T_i}\zeta_i e^{-\zeta_i^2} \right).
\end{align}
Had we omitted the imaginary part of the large-argument and small-argument approximations \cref{eq:large_argument,eq:small_argument}, we would have arrived at the dispersion relation $\varepsilon_r(\omega, k)=0$, which only has two real roots:
\begin{align}
\omega_r=\pm k\frac{c_s}{\sqrt{1+k^2\lambda_{De}^2}},
    \label{eq:ionac_real_roots}
\end{align}
where $c_s=\sqrt{k_BT_e/m_i}$ is the ion acoustic speed, and $k_B$ is Boltzmann's constant.

As we will see, the ion acoustic instability has a weak growth compared to the frequency, $\gamma\ll|\omega_r|$. The roots given in \cref{eq:ionac_real_roots} can therefore be taken as a good approximation to the real part of the roots of the full, complex dispersion relation. Further on, for weak instabilities, the imaginary part $\gamma$ can be found using the following general equation for weak instabilities \cite{treumann_advanced}:
\begin{align}
    \gamma = -\frac{\varepsilon_i(\omega_r, k)}
    {\left.\partial\varepsilon_r(\omega, k)/\partial\omega\right|_{\omega=\omega_r}}
\end{align}
After some algebraic manipulations, we arrive at the growth rate
\begin{align}
    \frac{\gamma}{|\omega_r|} &= \sqrt{\frac{\pi}{8}} \frac{1}{(1+k^2\lambda_{De}^2)^\frac{3}{2}} \times \nonumber\\
    &\left[\underbracket{\sqrt\frac{m_e}{m_i} \left(\frac{k u}{\omega_r}-1\right) e^{-\zeta_e^2}}_{\text{electrons}}
    \underbracket{-\left(\frac{T_e}{T_i}\right)^\frac{3}{2} e^{-\zeta_i^2}}_{\text{ions}} \right],
    \label{eq:ionac_growth}
\end{align}
where, due to the small-argument approximation, $e^{-\zeta_e^2}\approx 1$.

\Cref{eq:ionac_real_roots,eq:ionac_growth} are plotted in \cref{fig:disprel_ionac} in black for the parameters in \cref{tab:parameters}. For the ion acoustic case, it is convenient to normalize the axes by the \emph{ion} plasma frequency since the waves occur at ion scales, but we remark that this is still in accordance with the dimensionless groups in \cref{eq:groups}, since the mass ratio is also a group. Similar to in the Buneman case, the real part of the roots has asymptotes $\omega=\pm\omega_{pi}$ for large $k$. For short $k$, the two real roots follow the asymptotes $\omega=\pm c_s k$, which is a similar dispersion relation as sound waves of speed $c_s$ in neutral media. Only the positive branch has phase velocities $\omega/k$ that satisfy the Penrose criterion for instability, so the growth rate given by \cref{eq:ionac_growth} belong the the branch with positive real part in \cref{fig:disprel_ionac}. The fastest-growing mode's characteristics, as obtained from the plotting script, is listed in \cref{tab:ionac}.

\begin{table}
    \caption{Characteristics of the fastest-growing wave for the ion acoustic instability}
    \begin{tabular}{lrrr}
        \hline
        & $\frac{ku}{\omega_{pi}}$ & $\frac{\omega_r}{\omega_{pi}}$ & $\frac{\gamma}{\omega_{pi}}$ \\
        \hline
        Approximate kinetic theory                       & 2.01 & 0.627 & 0.041 \\
        Kinetic theory                                   & 1.94 & 0.640 & 0.038 \\
        \hline
        Simulation                                       & 1.94 & 0.383 & $-$0.001 \\
        \hline
Drift-averaged method                            &      &       & 0.018 \\
        \hline
    \end{tabular}
    \label{tab:ionac}
\end{table}

Let's turn our attention to \emph{when} these intabilities occur. The ion damping is largely determined by
\begin{align}
    \zeta_i^2 = \frac{T_e/T_i}{1+k^2\lambda_{De}^2}.
\end{align}
When $\zeta_i^2\gg 1$, or equivalently,
\begin{align}
    T_e\gg(1+k^2\lambda_{De}^2)T_i\geq T_i,
    \label{eq:ionac_crit_temp}
\end{align}
the ion damping term in \cref{eq:ionac_growth} can be neglected, and instability then occur when the electron term is positive, i.e., when
\begin{align}
    u>\frac{\omega_r}{k}=\frac{c_s}{\sqrt{1+k^2\lambda_{De}^2}}.
    \label{eq:ionac_crit_speed}
\end{align}
The ion damping term quickly gets stronger when $\zeta_i^2$ gets smaller, which explain why ion acoustic instabilities usually occur for $k\lambda_{De}\ll 1$. When this is the case, \cref{eq:ionac_crit_temp,eq:ionac_crit_speed} simplify to $T_e\gg T_i$, and $u>c_s$. It is still possible to get unstable ion acoustic modes when $T_e\sim T_i$, but then the drift speed must be even larger to compensate for the ion damping.

The ion acoustic case selected for this paper, and which is listed in \cref{tab:parameters}, has $T_e/T_i=64$, and $u/c_s=2.5$, and is therefore well within the range where ion acoustic instabilities should occur. To obtain growth rates \emph{without} the large-argument and small-argument approximations, we again resort to using the secant method on \cref{eq:disprel_asymmetric_kinetic}. We use the approximate solutions as an initial guess, and for each $k$, we plot the result in gray in \cref{fig:disprel_ionac}. We also included the growth rate of the branch with negative real part, and as is to be expected from the Penrose criterion, it is always negative. The fastest-growing mode (without the approximations) is also obtained from the plotting script, and listed in \cref{tab:ionac}.
  \section{Current-driven Langmuir Oscillations}\label{sec:langmuir}
When considering electrostatic phenomena, one usually obtains the electric field by solving the Poisson equation. For the symmetric and current-free two-stream case (\cref{fig:config_symmetric}), the charge density $\rho=0$ before any perturbation sets in. One might therefore erroneously conclude that the electric field $\vec E=\vec 0$. If, instead, we use Ampére's equation, we find an electric field oscillating at the plasma frequency. Although this has been shown before for cold electrons and immobile ions \cite{baumgaertel,sauer2015,sauer2016}, we here generalize it to arbitrary velocity distributions and multiple species, and show that oscillations occur at the plasma frequency for the conditions supposed to lead to the Buneman and ion acoustic instability.

Since all fields are uniform in the unperturbed state, $\nabla\rightarrow 0$, and Ampére's equation reduces to
\begin{align}
    \vec J=-\varepsilon_0\dot{\vec E},
    \label{eq:current_balance}
\end{align}
where the dot indicates the time derivative (the partial and total time derivatives coincide when $\nabla\rightarrow 0$). The current density $\vec J$ can also be written
\begin{align}
    \vec J=\sum\limits_s q_s n_s\vec u_s,
\end{align}
where $q_s$ is the charge of species $s$, and the density $n_s$ and bulk velocity $\vec u_s$ is defined from the zeroth and first order velocity moments of the distribution function $f_s$ in the usual way \cite{pecseli}.
Combining the above two equations, we arrive at
\begin{align}
    \varepsilon_0\dot{\vec E}=-\sum\limits_s q_s n_s \vec u_s,
    \label{eq:Eu-relation}
\end{align}

To eliminate $\vec u_s$, we need the equations of motion for each species $s$, and to arrive at appropriate equations for arbitrary velocity distributions, we have to start from the Vlasov equations. When taking velocity moments of the Vlasov equation, one normally ends up with the so-called BBGKY hierarchy, where the equation from the zeroth moment couple to the equation for the first moment, which again couple to the equation for the second moment and so on \cite{pecseli}. Solving this hierarchy of moments is usually no simpler than solving the Vlasov equations directly, unless the hierarchy is somehow truncated, as is the case for fluid approximations. Interestingly, since $\nabla\rightarrow 0$ in our case, this coupling disappears entirely. When $\nabla\rightarrow0$, the zeroth and first order velocity moments of the Vlasov equation for species $s$ are simple continuity and momentum equations:
\begin{align}
    & \dot n_s = 0, \label{eq:langmuir_continuity}\\
    & m_s\dot{\vec u}_s=q_s\vec E, \label{eq:langmuir_momentum}
\end{align}
where $m_s$ is the mass of species $s$. The implication is that the above two equations are valid not only in the cold fluid approximation, but for any velocity distribution, as long as the plasma is spatially uniform.

Taking the time-derivative of both sides of \cref{eq:Eu-relation} and using the equations of motion, we get a harmonic equation for the electric field:
\begin{align}
    \ddot{\vec E}+\omega_0^2\vec E=\vec 0,
\end{align}
where
\begin{align}
    \omega_0^2 = \sum\limits_s \omega_{ps}^2,
\end{align}
and $\omega_{ps}=\sqrt{q_s^2n_s/\varepsilon_0 m_s}$ is the plasma frequency of species $s$. The electric field can thus be written,
\begin{align}
    \vec E(t) = \Re{\vec A e^{-i\omega_0 t}},
    \label{eq:E_oscillation}
\end{align}
where $\vec A$ is some complex amplitude. For a single (mobile) species, $\vec u_s$ can be calculated very easily from \cref{eq:Eu-relation}, which then reduces to $\vec u_s=-\varepsilon_0 \dot{\vec E}/q_s n_s$. For multiple species, however, we must integrate \cref{eq:langmuir_momentum}:
\begin{align}
    \vec u_s(t)
&= \vec u_s(0) + \frac{q_s}{m_s\omega_0}
    \left(\Im{\vec A} - \Im{\vec A e^{-i\omega_0 t}} \right)
    \label{eq:u_oscillation}
\end{align}
The real and imaginary parts of $\vec A$ are given by initial conditions in $\vec E$ and $\vec u_s$, respectively. To determine the real part, set $t=0$ in \cref{eq:E_oscillation}. For the imaginary part, substitute \cref{eq:E_oscillation,eq:u_oscillation} into \cref{eq:Eu-relation}, and set $t=0$:
\begin{align}
    \Re{\vec A} &= \vec E(0) \label{eq:re_A} \\
    \Im{\vec A} &= -\frac{1}{\varepsilon_0 \omega_0}\sum\limits_s q_s n_s \vec u_s(0) \label{eq:im_A}
\end{align}
In the following subsections we shall consider some special cases of particular relevance to us.

\subsection{Electron-only oscillations}
\label{sec:eo_oscillations}
For both the Buneman instability (\cref{fig:config_asymmetric}) and the ion acoustic instability (\cref{fig:config_ionac}), we consider an initial configuration where ions have no drift velocity, and give rise to no current. If we consider the ions infinitely more massive than the electrons, they will remain at zero drift velocity, and we need only account for a single electron species, $s\in\{e\}$.

The frequency of oscillation in this case will simply be the electron plasma frequency,
and \cref{eq:u_oscillation,eq:re_A,eq:im_A} simplify to
\begin{align}
\vec u_e(t)=\Re{\left(\vec u_e(0)+i\frac{q_e}{m_e\omega_{pe}}\vec E(0)\right)e^{-i\omega_{pe}t}}.
    \label{eq:simple_oscillating_u}
\end{align}
If, in addition, $\vec E(0)=\vec 0$, then $\vec u_e(t)=\vec u_e(0)\cos(\omega_{pe}t)$.

Note that these oscillations are not just the small perturbations predicted at the plasma frequency by linearizing the Vlasov-Maxwell equations. Instead, we have shown from the \emph{non-linearized} equations that the entire electron velocity distribution, as depicted in \cref{fig:config_asymmetric}, oscillates back and forth around zero velocity with an amplitude equal to the initial drift velocity (if $\vec E(0)=\vec 0$) or larger!
Naturally, this causes the \emph{position} of the electrons to oscillate also, in much the same way as for Langmuir oscillations driven by perturbations in the charge density. In this case, however, we have not yet introduced any perturbation, and the charge density remains uniformly zero during the entire oscillation period (c.f., \cref{eq:langmuir_continuity}). The oscillations are purely current-driven. Since the oscillation frequency is so much larger than the growth rates for the ion acoustic instability and even the Buneman instability, it is reasonable that these oscillations alter these instabilities.

The current-driven Langmuir oscillations can also be understood in a qualitative manner (omitting vector notation for the sake of argument). As long as the electron velocity $u_e>0$, the current $J=q_en_eu_e<0$ ($q_e$ is negative), and for the Maxwell displacement current $\varepsilon_0\dot E$ to balance the current $J$ (see \cref{eq:current_balance}), the electric field must increase. Regardless of the initial electric field, $E$ will eventually become positive, and at this point the electrons have negative acceleration, and the electron peak in \cref{fig:config_asymmetric} moves leftwards. It cannot start moving rightward again until $E<0$, and this can only happen when $J=q_e n_e u_e$ has been larger than zero for some time, i.e., well after the electron peak has passed zero velocity.

\subsection{Electron--electron oscillations}
\label{sec:ee_oscillations}
For the two-stream instability, we consider two different initial configurations in one dimension. In one case we are in a frame of reference where the electron species have equal but oppsite velocities as illustrated in \cref{fig:config_symmetric}. The currents from the two species cancel, such that $J=0$. Consequently, the electric field remains constant, as can be seen from \cref{eq:current_balance}, and assuming it is initially zero, the drift velocities also remain constant (\cref{eq:langmuir_momentum}). Without a current, there can be no current-driven Langmuir oscillations.

Moving to the asymmetric reference frame depicted in \cref{fig:config_asymmetric} lead to a more curious result. Let both species have charge $q$, mass $m$ and density $n$. Species 2 has zero initial velocity, $u_2(0)=0$, while species 1 has initial velocity $u_1(0)=u_0$. The initial electric field is also zero. Using \cref{eq:re_A,eq:im_A} and writing out \cref{eq:u_oscillation} for both species yields,
\begin{align}
    u_1(t) = +\frac{1}{2}u_0 + \frac{1}{2}u_0\cos (\omega_0 t), \\
    u_2(t) = -\frac{1}{2}u_0 + \frac{1}{2}u_0\cos (\omega_0 t),
    \label{eq:ee_oscillations}
\end{align}
where $\omega_0=\sqrt{2}\omega_{pe}$.

Both species oscillate back and forth in velocity space, but remain in-phase, with the peaks always a velocity $u_0$ apart.
That the peaks remains at a fixed distance apart in velocity space can also be understood in another way: The two electron species can be considered a single species with a two-humped velocity distribution. Since the Langmuir oscillations only affect the drift velocity and not the distribution, the two humps must remain a distance $u_0$ apart.

A curious part of this result is that the electrons appear to be accelerated in the frame with a current, but not in the current-free frame, and yet the two reference frames are only separated by a constant velocity.

\subsection{Ion--electron oscillations}
\label{sec:ie_oscillations}
Let us again consider the Buneman and ion acoustic cases, but now with a finite ion mass. If the electrons have an initial speed $u_0$, the ions zero initial speed and the electric field is also initially zero, we find the solutions
\begin{align}
    u_e(t)=\frac{m_e}{m_i+m_e}u_0 + \frac{m_i}{m_i+m_e}u_0\cos(\omega_0 t), \\
    u_i(t)=\frac{m_e}{m_i+m_e}u_0 - \frac{m_e}{m_i+m_e}u_0\cos(\omega_0 t),
\end{align}
where $\omega_0=\sqrt{1+(m_e/m_i)}\omega_{pe}$.

Contrary to for the electron-only oscillations, where the ions had infinite mass, we now get oscillations of both species about a common velocity $m_eu_0/(m_i+m_e)$, but in opposite phase. For $m_i\gg m_e$, the ions oscillate with a much smaller amplitude than the electrons, but we remark that the electrons also have a slightly diminished amplitude compared to in the electron-only oscillations. Interestingly, the relative drift velocity remains the same as for the electron-only oscillations,
\begin{align}
    u(t) = u_e(t)-u_i(t)=u_0\cos(\omega_0 t),
    \label{eq:ie_oscillations_relative}
\end{align}
except for the slightly increased frequency $\omega_0$.

\subsection{Gyrations in velocity space}
Finally, let us consider a more esoteric example. It is possible to have a single (mobile) species, whose distribution follow a circular trajectory of radius $u_0$ in 2D velocity space:
\begin{align}
    \vec u(t)=u_0
    \begin{bmatrix}
\cos(\omega_0 t) &
        \sin(\omega_0 t)
    \end{bmatrix}^T.
\end{align}
This is \cref{eq:simple_oscillating_u} when
$\vec E(0) = [\begin{matrix}0 & u_0\omega_0m_e/q_e\end{matrix}]^T$.

Beware that since the oscillations do not affect the shape of the distribution, the distribution itself do not rotate. If, for instance, the initial distribution makes a square-shaped contour in the velocity space, the square would not rotate, but follow a circular trajectory.

Interestingly, there is no damping in the amplitude of these current-driven Langmuir oscillations, even if we account for kinetic theory, which usually lead to Landau damping. In a real scenario, however, perfect uniformity do not exist, and when $\nabla$ is not exactly zero, it is conceivable with some Landau damping.
 \section{Revisiting the Electrostatic Approximation} \label{sec:helmholtz}

It may appear a paradox that these current-driven Langmuir oscillations are
electrostatic phenomena, but seemingly cannot be explained by the Poisson
equation. To resolve this paradox, we need to look more carefully at the
electrostatic approximation.

In electromagnetism, the curl and divergence of the $\vec E$-field is given by
Faraday's law and Gauss' law \cite{griffiths}:
\begin{align}
    & \curl\vec E = -\pder{\vec B}{t}, \label{eq:faraday}\\
    & \div\vec E = \frac{\rho}{\varepsilon_0}, \label{eq:gauss}
\end{align}
where $\rho$ is the charge density and $\vec B$ the magnetic flux density.
According to Helmholtz' theorem \cite{griffiths,arfken,cheng}, for any prescribed values of the curl and divergence (i.e., the right hand sides), there \emph{exists} a vector field $\vec E$.
It is easy to forget that the above is not enough for $\vec E$ to be \emph{unique}.
Indeed, any vector field which is both curl- and divergence-free (a \emph{Laplacian vector field}) can be added to $\vec E$, and the result would still be a solution of the above equations. 
For $\vec E$ to be unique, its normal component must be prescribed on the boundary, or, for unbounded systems, $\vec E$ must tend to zero as the distance $r\rightarrow\infty$. This is often a reasonable criterion. Consider for instance the electric field due to some source. For the field energy $\propto\int E^2\,\dd \vec x$ to be finite, the electric field magnitude $E$ must decay sufficiently fast towards zero as $r\rightarrow\infty$.

However, for the current-driven Langmuir oscillations, the $\vec E$-field do \emph{not} decay, but is instead uniform, and given by \cref{eq:E_oscillation}. Because of the electrostatic approximation, and since $\rho=0$, Faraday's and Gauss' law simplify to
\begin{align}
    & \curl\vec E = \vec 0, \label{eq:curl-free_E}\\
    & \div\vec E = 0. \label{eq:div-free_E}
\end{align}
Indeed, \cref{eq:E_oscillation} is readily seen to satisfy these for \emph{any} time instant.
Of course, in practice the oscillations cannot extend indefinitely and have infinite energy, but as often in theory, what we consider is an idealized case ($\nabla\rightarrow 0$).

Next, a related theorem, also often referred to as the Helmholtz' theorem \cite{griffiths,arfken,cheng}, states that any twice continuously differentiable field $\vec E$ can be expressed using a curl-free and a divergence-free component:
\begin{align}
    \vec E=-\nabla\phi+\curl\vec F.
\end{align}
This is valid even for non-decaying $\vec E$-fields. Moreover, since $\vec E$ is curl-free in the electrostatic regime, we can omit the second term, and express $\vec E$ using only the curl-free part $-\nabla\phi$. We emphasize, however, that this curl-free term may also be divergence-free. Indeed, the non-decaying, curl- and divergence-free electric field in \cref{eq:E_oscillation}, can be expressed through a potential
\begin{align}
    \phi(t)=\vec E(t)\cdot\vec x, \label{eq:consistent_potential}
\end{align}
where $\vec x$ is the posistion vector. Substituting $\vec E=-\nabla\phi$ into Gauss' law gives us the Poisson equation, as usual:
\begin{align}
    -\nabla^2\phi = \frac{\rho}{\varepsilon_0}. \label{eq:poisson}
\end{align}

To conclude, the Poisson equation is a valid description for the current-driven Langmuir oscillations after all. The failure of capturing current-driven Langmuir oscillations in electrostatic simulations and analytical derivations based on the Poisson equation is instead due to incorrect boundary conditions. Since we know the solution analytically, we can work out the correct boundaries. If we were to solve \cref{eq:gauss,eq:curl-free_E} numerically, we would need uniform, but time-varying Dirichlet boundaries on the $\vec E$ field as given by \cref{eq:E_oscillation}. If instead we were solving the Poisson equation, we would need \emph{non-uniform}, time-varying Dirichlet boundaries, according to \cref{eq:consistent_potential}.

In numerical studies of instabilities, where a spatial periodicity is given by the wavenumber, it seems reasonable to use periodic boundary conditions. However, periodic boundaries, together with \cref{eq:gauss,eq:curl-free_E}, do not provide a unique $\vec E$. Indeed, add any uniform field, and you get another solution of the equations that are also periodic. Solving the \emph{Poisson} equation and enforcing periodicity in $\phi$ would be even worse, since, as we now know, the potential in \cref{eq:consistent_potential} is not periodic (and besides, the Poisson equation do not have a unique solution to begin with when all boundaries are periodic). Admittedly, it is difficult to know what the correct boundary conditions for the electrostatic equations are, once we start including instabilities or other phenomena beyond just the Langmuir oscillations. However, when solving the full Maxwell set of equations (or often just the curl equations), periodic boundary conditions for all electromagnetic fields is adequate. Together with initial conditions satisfying Maxwell's divergence equations, a unique and correct solution can be obtained.
 \section{Numerical results}
\label{sec:simulations}
Let us now turn to numerical simulations of the configurations depcited in \cref{fig:config}, in order to see how the instabilities are modified by current-driven Langmuir oscillations. To this end, we use the openly available state-of-the-art code Gkeyll 2.0, which is developed at PPPL (see for instance \citet{gkeyll,hakim,juno,cagas}). Of utmost importance to us is the fact that Gkeyll solves the full set of Vlasov-Maxwell's equations, instead of relying upon the Poisson equation, since, as we have seen in the preceeding section, current-driven Langmuir oscillations can hardly be predicted from the Poisson equation. Moreover, the discontinuous Galerkin discretization employed in Gkeyll is exactly energy conserving \cite{juno}. Since we're solving the Vlasov equation directly, there is also no particle noise in our simulations, as opposed to in particle-in-cell simulations. All our simulations have one spatial dimension and one velocity dimension (1X1V), both of which are uniformly discretized.

Periodic boundary conditions are used for the spatial axis, and because of the finite length of the domain, only wave modes where an integer number of wavelengths fit inside the simulation domain can be captured by the simulations \cite{shalaby}. Thus, to study the growth rate of the fastest-growing modes, we choose the domain length to be one wavelength, $2\pi/k$, for the $k$ listed in \cref{tab:twostream,tab:buneman,tab:ionac}, with a spatial resolution of 0.02 wavelengths. Since the simulations are energy-conserving and particle noise-free, they do not produce any growing modes unless given an initial perturbation (at least not within a reasonable time). We do this by setting the inital distributions to
\begin{align}
    f_{10}(x,v,0) &= nf(v-u_{10}; v_{\mathrm{th},1}), \\
    f_{20}(x,v,0) &= n(1+A\cos(kx))f(v-u_{20}; v_{\mathrm{th},2}),
\end{align}
where $f(v; v_{\mathrm{th},s})$ is the Maxwellian/normal distribution with zero mean and standard deviation of $v_{\mathrm{th}}$, $n$ is the unperturbed plasma density, and $k$ is again the wavenumber of the fastest-growing mode from the tables. We take $A=10^{-3}$ for the two-stream cases, but for the more slowly-growing Buneman and ion acoustic cases we set $A=10^{-2}$ for faster onset of the instability. One might also have added perturbations for all other modes supported by the discretization of the domain, but in our case we are interested only in this one mode. The wavenumber $k$ is thus an \emph{input} to the simulation. In order for the simulations to be consistent, the intial conditions should also satisfy Gauss' law, $\nabla\cdot\vec E=\rho/\varepsilon_0$. This is achieved by setting the initial electric field to
\begin{align}
    E(x,0)=\frac{nq_2A}{k}\sin(kx),
\end{align}
where $q_2$ is the charge of species 2. The magnetic flux density is initialized to zero. The periodic boundary conditions also effectively prevent any convective modes, where $k$ is complex, since that would require different values on the left and right boundaries.

Gkeyll enforces zero particle flux as boundary conditions at the upper and lower velocities in phase space, in order to prevent loss of energy or momentum \cite{juno}. It is important, however, that the upper and lower velocity limits are sufficiently far away from the bulk of the distribution functions, lest numerical instabilities arise at the boundary, and the program crashes. An adequate domain size for the electrons is found to be $[-4u_0, 4u_0]$ for the two-stream and Buneman cases, where $u_0$ is the initial drift velocity, and $[-8v_{\mathrm{th},e}, 8v_{\mathrm{th},e}]$ for the ion acoustic case. For the ions, the domain $[-6v_{\mathrm{th},i}, 6v_{\mathrm{th},i}]$ is used for all simulations. The velocity space resolutions are 0.2 thermal speeds for both species. Since we report our results in non-dimensional groups, the exact physical parameters used in the simulations do not really matter, but for the sake of transparency, we set $n=1$, the electron mass and charge to $1$ and $-1$, respectively, and $\varepsilon_0=\mu_0=1$. This makes the speed of light one, and thus all velocities should be less than one. We achieve this by setting $u_0=0.02$ for the ion acoustic case, and $u_0=0.1$ for the other cases. The rest of the parameters can be inferred from \cref{tab:parameters}. For the sake of reproduceability, all our Gkeyll input files are available in the supplementary material \cite{supplementary}.

\subsection{The current-free two-stream case}
\begin{figure*}
    \centering
    \subfloat[The current-free two-stream case (\cref{fig:config_symmetric})]{
        \includegraphics{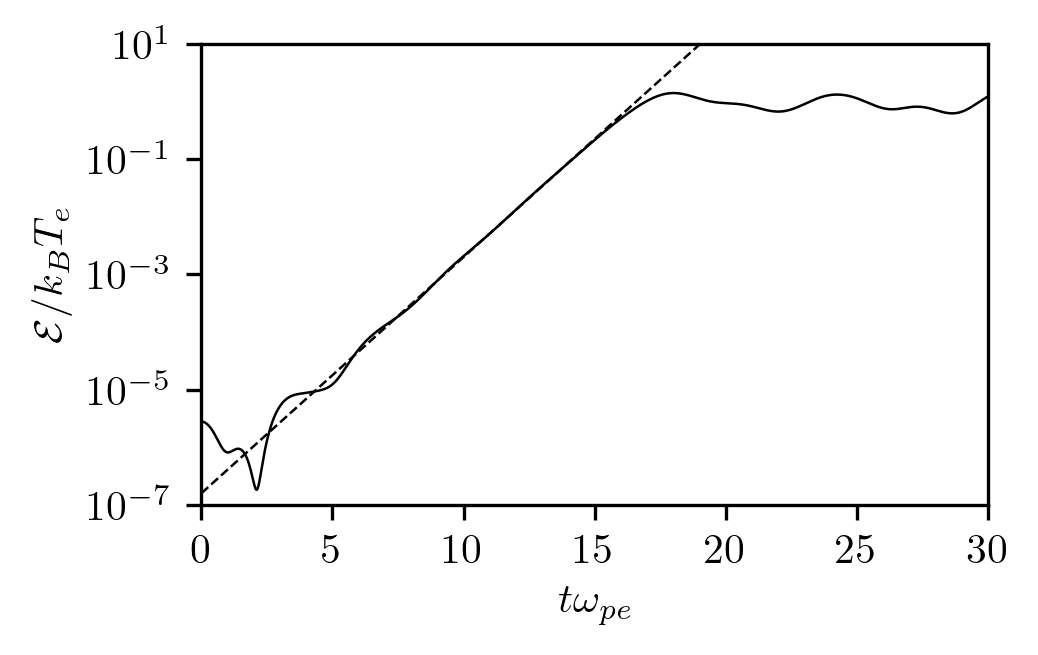}
        \label{fig:twostream_current_free}
    }
    \subfloat[The two-stream case (with current, \cref{fig:config_asymmetric})]{
        \includegraphics{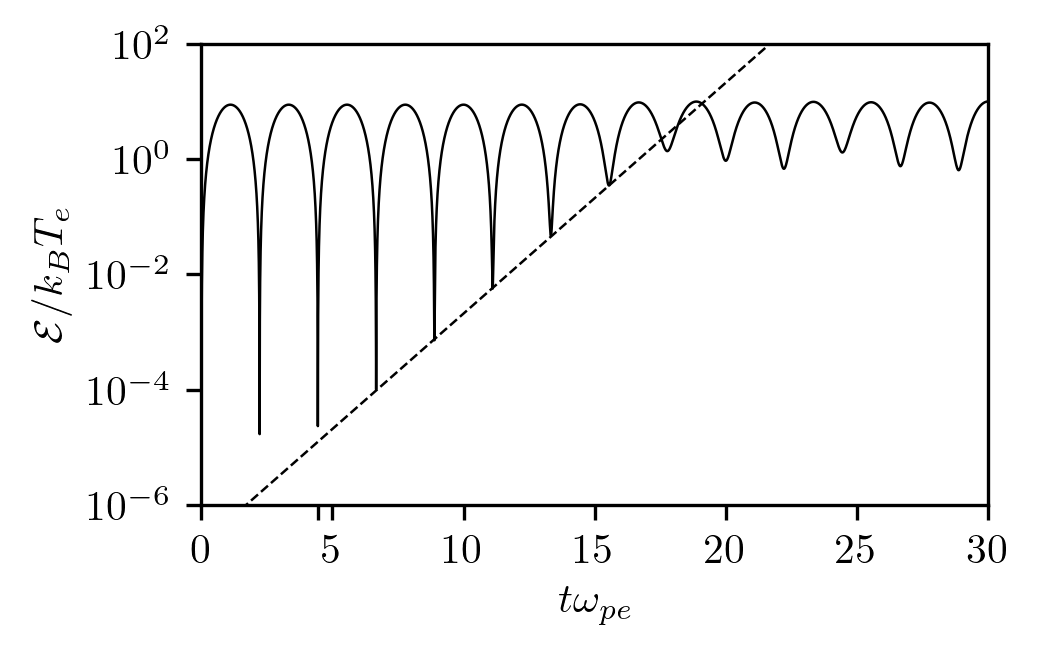}
        \label{fig:twostream_current}
    }

    \subfloat[The Buneman case (\cref{fig:config_asymmetric})]{
        \includegraphics{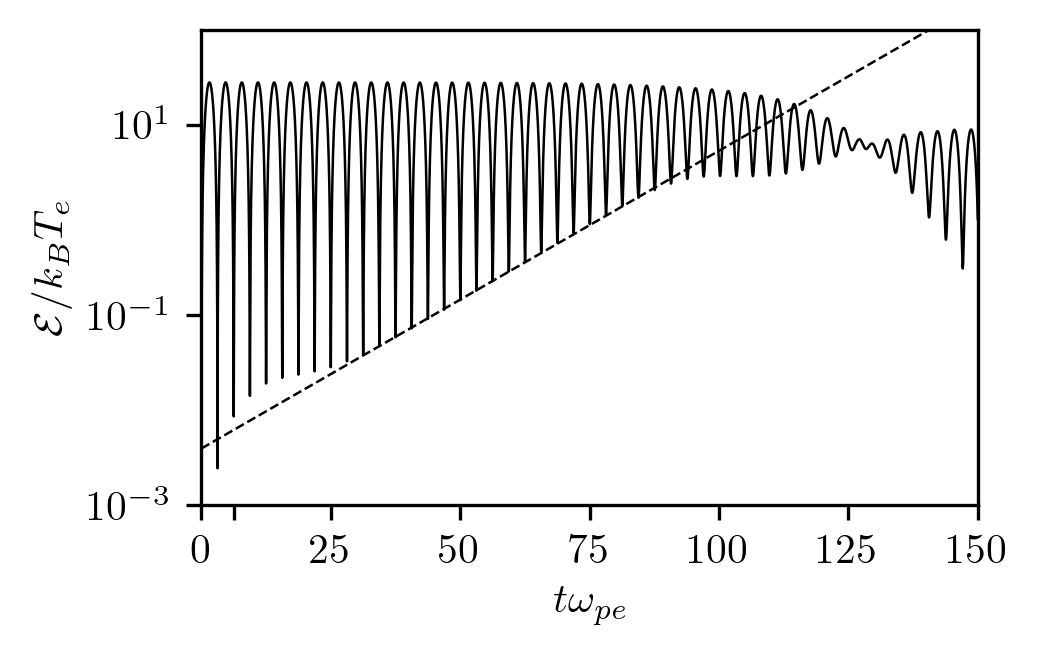}
        \label{fig:buneman}
    }
    \subfloat[The Buneman case but without initial perturbation]{
        \includegraphics{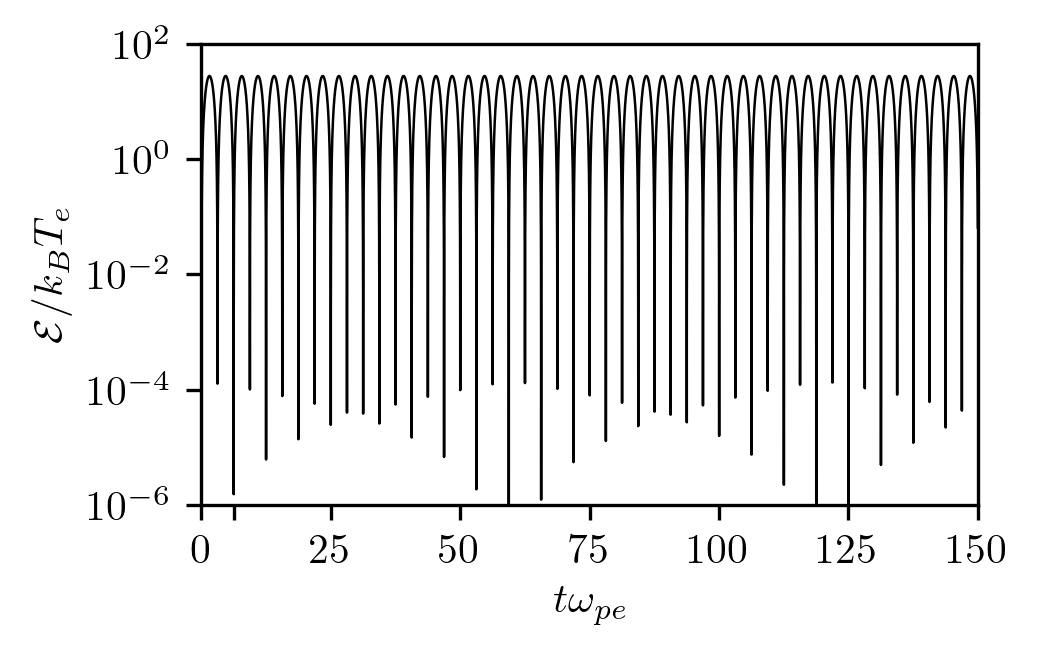}
        \label{fig:buneman_nopert}
    }

    \subfloat[The ion acoustic case (\cref{fig:config_ionac})]{
        \includegraphics{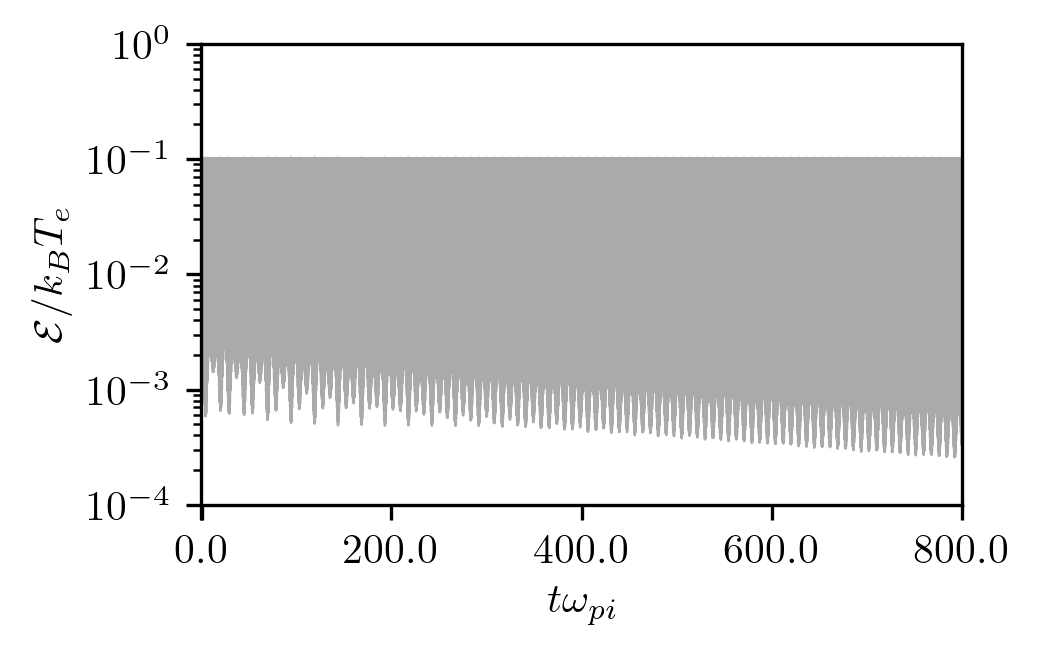}
        \label{fig:ionac}
    }
    \subfloat[The ion acoustic case (zoomed in)]{
        \includegraphics{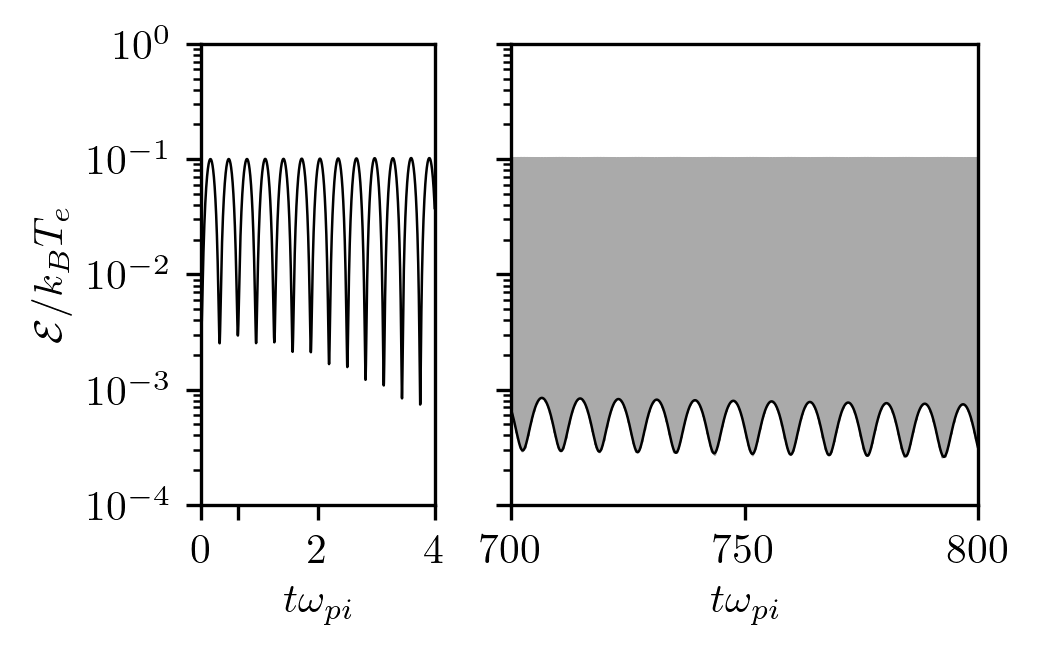}
        \label{fig:ionac_broken}
    }
    \caption{Evolution of electric field energy in the simulations. In \cref{fig:twostream_current_free} a slope is fitted to the energy to determine the growth rate, whereas in \cref{fig:twostream_current,fig:buneman}, a slope has been fitted to the \emph{minima} of the energy (dashed). For the ion acoustic case the duration is so long that the oscillations appear as a solid, gray area when not zoomed in sufficiently. A dampened sinusoid has been fitted to the minima in the right part of \cref{fig:ionac_broken} (black) to obtain the growth rate and frequecy. In \cref{fig:twostream_current,fig:buneman,fig:buneman_nopert,fig:ionac_broken}, a separate tick mark has been placed on the $x$-axis where the theory in \cref{sec:langmuir} predicts the first period of Langmuir oscillations to end.}
    \label{fig:energy}
\end{figure*}

As a benchmark case, we first present results for the current-free two-stream case. Snapshots of the phase space is shown in the left column of \cref{fig:ph_twostream}, whereas a full animation is available in the supplementary material
\cite{supplementary}. The evolution of the electric field energy,
\begin{align}
    \mathcal E=\frac{1}{2}\varepsilon_0\int E^2 \,\dd x,
    \label{eq:E_energy}
\end{align}
is plotted in \cref{fig:twostream_current_free}.
In the phase space, we initially see two Maxwellian-distributed electron species, in accordance with \cref{fig:config_symmetric}. During the linear regime, a spatially sinusoidal perturbation in drift velocity gradually increases.
At $t\omega_{pe}$ of about 15--17, nonlinearities sets in, as witnessed both from the non-sinusoidal shape in phase space in \cref{fig:ph_twostream}, as well as from the flattening of the energy in \cref{fig:twostream_current_free}. At $t\omega_{pe}=50$, we can see the formation of a phase space vortex so archetypical of the two-stream instability.

For the linear phase, $E\propto e^{\gamma t}$, and accordingly $\mathcal E\propto e^{2\gamma t}$. Therefore, to determine the growth rate $\gamma$, we fit a linear function with slope $2\gamma t$ to the logarithm of the energy in the window $t\omega_{pe}\in[8,15]$ (dashed line). The resulting growth rate is included in \cref{tab:twostream}, and it is only $0.2\%$ off the kinetic theory. Admittedly, the window $[8,15]$ is based on our judgment, and an inferior choice would lead to a slightly larger error. Nevertheless, it is clear that our simulations are quite capable of reproducing the correct behavior.

\begin{figure*}
    \centering
    \includegraphics{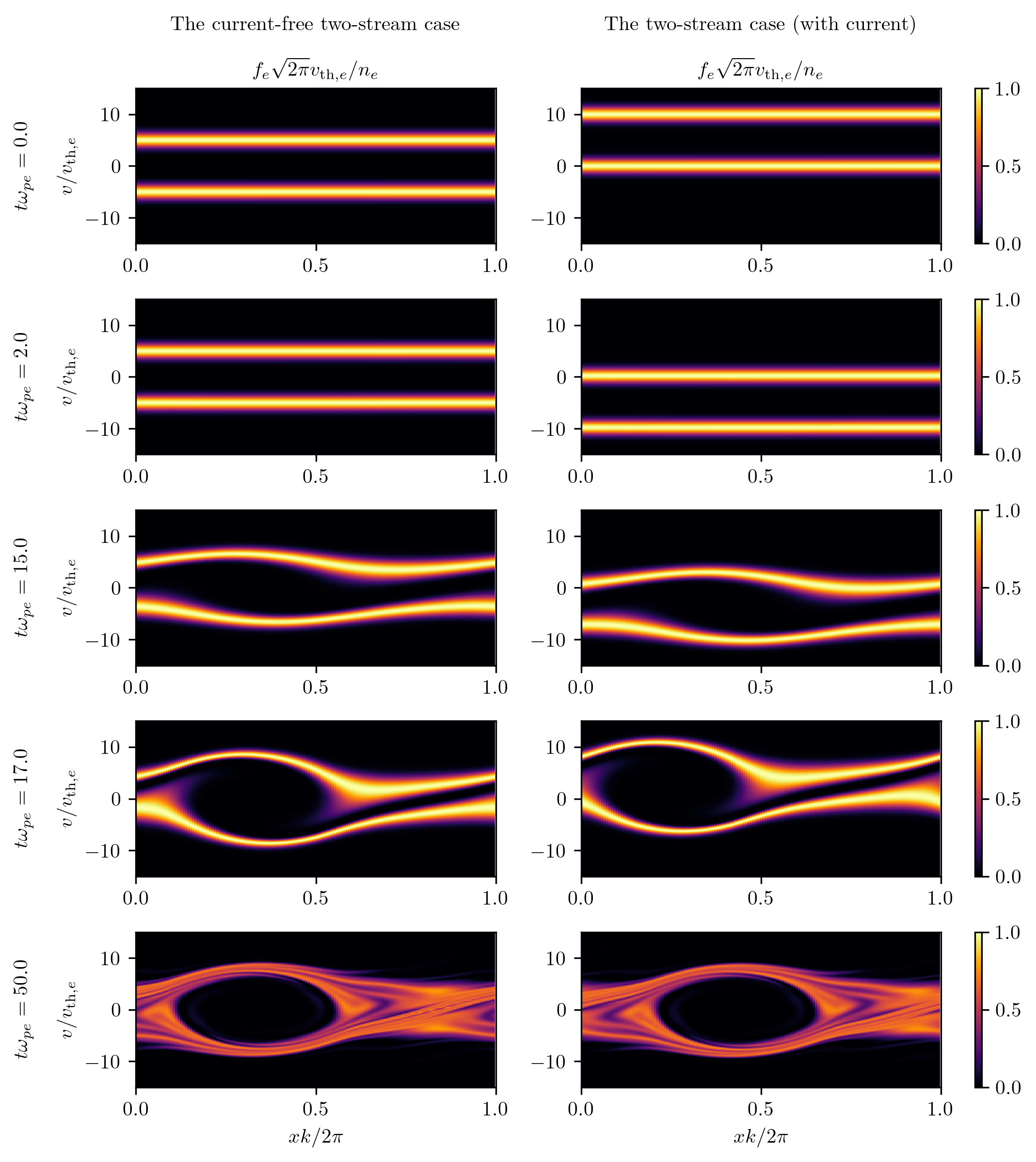}
    \caption{Snapshots of the electron distribution in phase space for the current-free two-stream case (left) and the two-stream case with current (right) at different time instants (indicated to the left). In the simulations, $f_e$ is the \emph{combined} distribution of both electron species, which is initially two-humped along the velocity axis. In the plots, the distribution is normalized to have peaks of unity in the initial state. The $x$-axis is normalized with respect to the wavelength $2\pi/k$ of the fastest-growing mode with wavenumber $k$, whereas the velocity axis is normalized with respect to the electron thermal speed. Best viewed in color.}
    \label{fig:ph_twostream}
\end{figure*}

It is interesting that there is a growth of electric field energy despite the fact that there is no source of energy in the simulations (periodic boundaries and energy conserving algorithms). However, since the species are initialized with different drift velocities, they are not at thermal equilibrium. This means that the relative drift energy is free energy that can feed an instability \cite{hasegawa}. Presumably, for a longer time-scale than we have simulated, the instability would thermalize the species, meaning that they combine into a single Maxwellian distribution with zero drift velocity, but with a higher thermal speed than the two species initially had, such that the total energy would be conserved. The phase space snapshot at $t\omega_{pe}=50$ already hints at this happening.

\subsection{The two-stream case}

Next, we turn our attention to the two-stream case \emph{with} current, i.e., initialized with the electron distributions placed asymetrically in velocity space as depicted in \cref{fig:config_asymmetric}. The initial phase space distribution depicted in \cref{fig:ph_twostream} (right column) is indeed in accordance with this, but already at $t\omega_{pe}=2$, the simulated distribution has changed dramatically! As is even more evident in the animation in the supplementary material \cite{supplementary}, the drift velocity of the two species oscillate \emph{in-phase}, between two extrema. This is in perfect agreement with the electron--electron oscillations predicted in \cref{sec:ee_oscillations}. According to to those predictions, the period of these oscillations should be $2\pi\omega_0^{-1}\approx 4.44\omega_{pe}^{-1}$, meaning that at $t\omega_{pe}=2$, almost half a period has elapsed. This agrees well with \cref{fig:ph_twostream}.

This simulation also show strong oscillations in the electric field energy (\cref{fig:twostream_current}).
The electric field naturally gains energy when the average drift velocity of the two species decrease, and vice versa, such that energy is conserved within the electric field and the motion of the species.
An oscillating field is also in accordance with \cref{eq:E_oscillation}, and the
period $2\pi\omega_0^{-1}$ is indicated with an extra tick mark on the $x$-axis in \cref{fig:twostream_current}. Recall that squared quantities like the energy oscillates at twice the frequency, such that two periods have elapsed at the extra tick mark (c.f., \cref{eq:E_energy}).

Curiously, the minima in the energy happen to be along a straight line in the region $t\omega_{pe}\in[5,17]$. We hypothesize that a two-stream instability grows in addition to the current-driven Langmuir oscillations, and that during the linear regime, the two are more or less uncoupled.
This is also consistent with the fact that the phase space snapshots of the two two-stream cases are similar, except for an oscillating up--down motion.
The total electric field energy will then be the energy of the oscillations, plus the energy of the instability. The minima occur when the oscillations have zero electric energy, and thus the energy at these points consist only of the unstable two-stream energy. If we fit a slope $2\gamma t$ through the minima in this region, we obtain a $\gamma$ that is $2.5\%$ off the kinetic theory for the two-stream instability (c.f. \cref{tab:twostream}). We consider $2.5\%$ to be a good agreement which support our hypothesis, especially considering that the estimate is based only on the five points within the window where there is a minimum.

\subsection{The Buneman case}
\begin{figure*}
    \centering
    \includegraphics{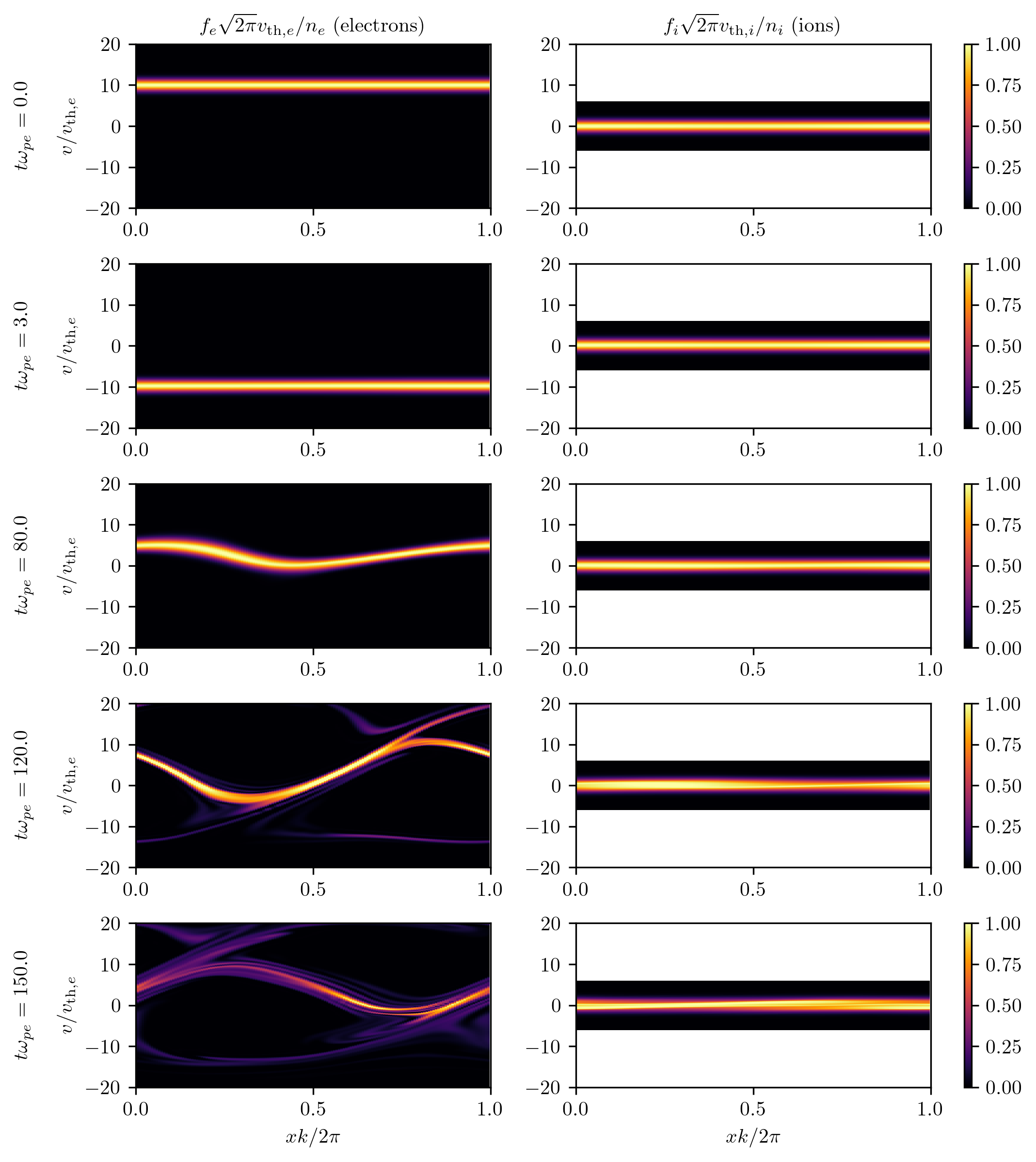}
    \caption{Snapshots of the electron (left) and ion (right) distribution in phase space for the Buneman case at different time instants (indicated to the left). For comparability, both distributions are plotted using the same axes, although the simulation domain is smaller along the velocity axis for the ions than the electrons. The white area in the ion distribution is the region outside the simulation domain. In the plots, the distributions are normalized to have peaks of unity in the initial state. The $x$-axis is normalized with respect to the wavelength $2\pi/k$ of the fastest-growing mode with wavenumber $k$, whereas the velocity axis is normalized with respect to the \emph{electron} thermal speed. Best viewed in color.}
    \label{fig:ph_buneman}
\end{figure*}

The energy for the simulation of the Buneman instability is plotted in \cref{fig:buneman}, whereas the phase space distribution is plotted separately for the electrons and ions in the left and right columns of \cref{fig:ph_buneman}, respectively (since the two species have different charge and mass, they can no longer be captured in a single distribution function).
Again, the initial phase space distributions is specified as in \cref{fig:config_asymmetric}, but at $t\omega_{pe}=3$, the electron drift velocity have moved from $10v_{\mathrm{th},e}$ to approximately $-10v_{\mathrm{th},e}$. Although it is difficult to see, the ion drift velocity has increased by a tiny amount during the same interval (animations available in the supplementary material \cite{supplementary}). This is consistent with the electron-only oscillations detailed in \cref{sec:eo_oscillations}, or more precisely, the ion--electron oscillations in \cref{sec:ie_oscillations}. The ion--electron oscillations have a period of $2\pi\omega_0^{-1}\approx 6.25\omega_{pe}^{-1}$, meaning that almost half a period has elapsed by $t\omega_{pe}^{-1}=3$. This is consistent with the phase space snapshots of the simulation. An extra tick mark in the energy plot, \cref{fig:buneman}, also indicate that the simulations have captured the ion--electron oscillations.

In this case also, it appears to be a linear slope in the energy minima in the interval $t\omega_{pe}\in[30,90]$. As can be seen from \cref{fig:ph_buneman}, the sinusoidal shape associated with a linear perturbation also starts deteriorating by the end of this interval. The growth rate obtained by fitting a linear slope to the minima in the interval, the dashed line in \cref{fig:buneman}, is barely a third of the growth rate predicted by the kinetic theory for the Buneman instability (c.f., \cref{tab:buneman}). In \cref{sec:modified}, we propose an explanation, and suggest a theoretical method which agree better with the simulations.

To remove any doubt that the upward slope in the minima is not merely due to numerical noise and inaccuracies, we include in \cref{fig:buneman_nopert} a plot of the energy when \emph{not} seeding the simulations with an initial perturbation, i.e., when $A=0$. Clearly, the minima are are much closer to zero in this case, and they remain low even after a moderately long time. Further on, we can report that even at $t\omega_{pe}=150$, the electron and ion densities remain uniform, and everywhere within $1\pm 2\cdot10^{-9}$ times the initial density. This agrees with the oscillations not being density-driven, and is also a testament to the quality of the Gkeyll code.

\subsection{The ion acoustic case}

The ion acoustic simulations also show oscillations consistent with ion--electron current-driven Langmuir oscillations. In the energy plot in \cref{fig:ionac} these oscillations are so rapid compared to the scale on the $x$-axis that the plotted lines appear like a solid, gray mass. The left part of \cref{fig:ionac_broken} show a small enough fraction of the simulation that the oscillations can again be recognized, and indeed, it coincides with the ion--electron oscillations as indicated by the extra tick mark at the period $2\pi\omega_0^{-1}\approx 0.63\omega_{pi}^{-1}$.

Ion acoustic instabilities are predicted to grow at a much slower rate than the other cases, and therefore need a much longer simulation time. For example, for the amplitude of the instability to grow by a couple orders of mangitude, or a factor $\sim e^{\gamma t}=e^5$, the simulation should run until $t\omega_{pi}\approx132$ for the growth rate predicted by the kinetic theory (c.f. \cref{tab:ionac}). Our simulation time is well beyond this and still no growth is visible. If anything, there appear to be a very weak damping, or downward slope in the energy. Close inspection reveal that the energy minima, i.e., the energy presumably due to the ion acoustic mode, this time appear to be on a sinusoidal curve. This is especially true in the later parts of the simulations, after some initial transients have died out. Since the ion acoustic instability has weak growth (or damping), i.e., $|\gamma|\ll|\omega_r|$, this actually makes sense. The linear theory predicts that the energy due to the ion acoutic mode is $E\propto e^{\gamma t}\cos(\omega_r t + \theta)$, and when $|\gamma|\ll|\omega_r|$, this expression can be interpreted as a sinusoid with slowly varying amplitude. The corresponding theoretical energy is then $\mathcal E\propto e^{2\gamma t}[1+\cos(2\omega_r t + 2\theta)]$. To determine the oscillation frequency and growth rate of the oscillations in the simulation, we fit the function $e^{2\gamma t}[c_1+c_2\cos(2\omega_r t + c_3)]$ to the minima of the energy in the window $t\omega_{pi}\in[700,800]$ (solid black line in \cref{fig:ionac_broken}). In addition to $\gamma$ and $\omega_r$, we have introduced the fitting coefficients $c_1$, $c_2$ and $c_3$, and ideally $c_1$ and $c_2$ should be equal. However, we get a better fit when we allow them to be different, in which case $c_2$ is about an order-of-magnitude smaller than $c_1$. We do not offer any conclusive explanation for this, but it might be because we are approaching the limit to how deep the minima can be in our simulation.

\begin{figure*}
    \centering
    \includegraphics{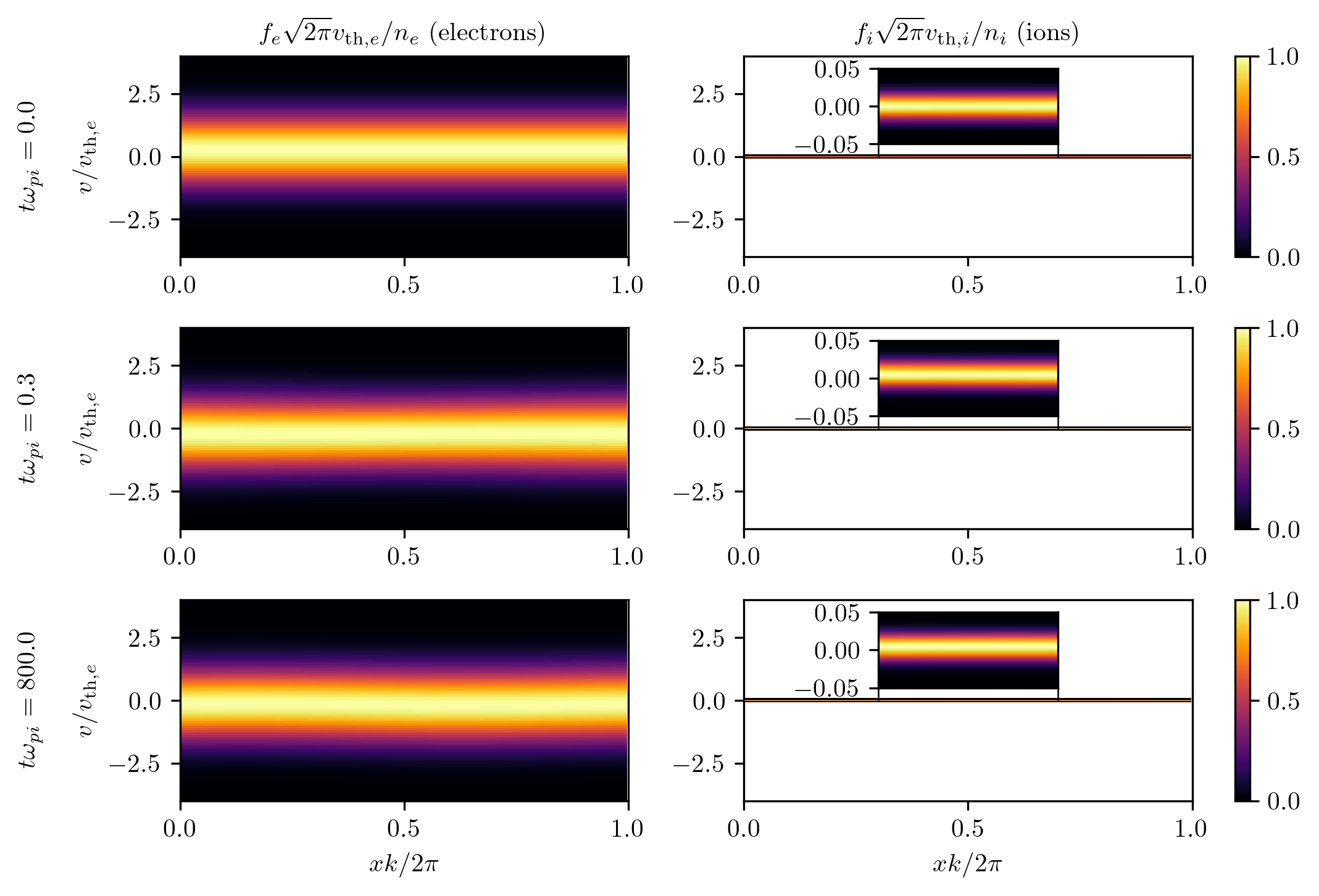}
    \caption{Snapshots of the electron (left) and ion (right) distribution in phase space for the ion acoustic case at different time instants (indicated to the left). For comparability, both distributions are plotted using the same axes, although the simulation domain is smaller along the velocity axis for the ions than the electrons. The white area in the ion distribution is the region outside the simulation domain. Since the ions have so much lower thermal speed than the electrons, the entire distribution is but a thin line. For a better view, a zoomed-up picture of the ion distribution is provided for the middle part in the inset axes (which share $x$-axis with the mother plot). In the plots, the distributions are normalized to have peaks of unity in the initial state. The $x$-axis is normalized with respect to the wavelength $2\pi/k$ of the fastest-growing mode with wavenumber $k$, whereas the velocity axis is normalized with respect to the \emph{electron} thermal speed. Best viewed in color.}
    \label{fig:ph_ionacoustic}
\end{figure*}

In any case, the results of the fitting procedure are given in \cref{tab:ionac}, and the first thing to observe is that the frequency of the oscillations in the minima appear to be of the right order-of-magnitude to be ion acoustic waves. Nonetheless, it is not a perfect match, being only approximately 60\% of the predicted frequency. Moreover, the growth has been replaced by a damping. We hypothesize that the current-driven Langmuir oscillations modify the ion acoustic frequency, and stabilizes the instability, but have currently no explanation for this beyond that.

Since the ion acoustic mode do not grow, it is too weak to be observed from the phase space snapshots in \cref{fig:ph_ionacoustic}. Nonetheless, the current-driven Langmuir oscillation can be seen by careful inspection, and at $t\omega_{pi}=0.3$ almost half a period has passed. Animations are available in the supplementary material \cite{supplementary}.
 \section{The drift-averaging method}\label{sec:modified}

As we have discussed, textbooks show that the configurations depicted in \cref{fig:config_asymmetric,fig:config_ionac} are unstable by assuming a \emph{constant} relative drift velocity $u$. Yet, we have seen that these configurations are subject to strong current-driven Langmuir oscillations in $u$, at a frequency large compared to the traditionally derived growth rates. This clearly affect the outcome. It might still be that the instabilities co-exist with the Langmuir oscillations in a form not too different from their usual form -- at least in the linear regime -- as hinted at by the minima in the energy in \cref{fig:energy}. Even so, the constant drift velocity assumption in the traditional derivations is questionable, at best.

A more correct approach would be to add a perturbation on top of an \emph{oscillating} equilibrium instead of a constant one.
More precisely, the \emph{unperturbed} state could have an oscillating velocity $u(t)$ and electric field $E(t)$ as given by the current-driven Langmuir oscillations described in \cref{sec:langmuir}, and perturbations $\delta u$ and $\delta E$ would be added on top of that, respectively. The unperturbed density can be assumed constant as usual. Moreover, since we deal with electrostatic phenomena, we can rely on the Vlasov-Poisson equations, as established in \cref{sec:helmholtz}, so long as we use the correct unperturbed state. One can then imagine substituting $u+\delta u$, and so forth, into the Vlasov-Poisson equations, eliminate the unperturbed quantities, and derive an equation for the perturbed state. The perturbed state could then be assumed proportional to $e^{i(kx-\omega t)}$ as usual, in the hope of arriving at a corrected dispersion relation. Unfortunately, this poses some difficulties: In a normal linearization procedure, one assumes the perturbations to be much smaller than the equilibrium, e.g., $\delta u\ll u(t)$. However, $u(t)=0$ twice per period, and in the vicinity of those zero-crossings, $\delta u$ will no longer be smaller than $u(t)$. Moreover, dispersion relations such as the one in \cref{eq:generic_disprel} usually assume $E(t)=0$, which is not true in the presence of current-driven Langmuir oscillations.

A simpler approach is motivated by the fact that the simulated growth rate for the two-stream case agrees well with the established theory, despite the oscillations in $u(t)$ and $E(t)$. What sets the two-stream case apart from the Buneman and ion acoustic cases, is that the \emph{relative} drift velocity between the two electron species actually \emph{do} remain constant (c.f. \cref{sec:ee_oscillations}). Contrariwise, for the Buneman and ion acoustic cases we have ion--electron oscillations, and the two distributions oscillate in opposite phases and cross one another (c.f. \cref{sec:ie_oscillations}).
Our approcah is therefore to use the standard dispersion relations, but to average the growth rate over varying relative drift velocities $u(t)=u_e(t)-u_i(t)$.

\subsection{The Buneman case}

\begin{figure*}
    \subfloat[Fluid, $u/u_0=1$]{\includegraphics{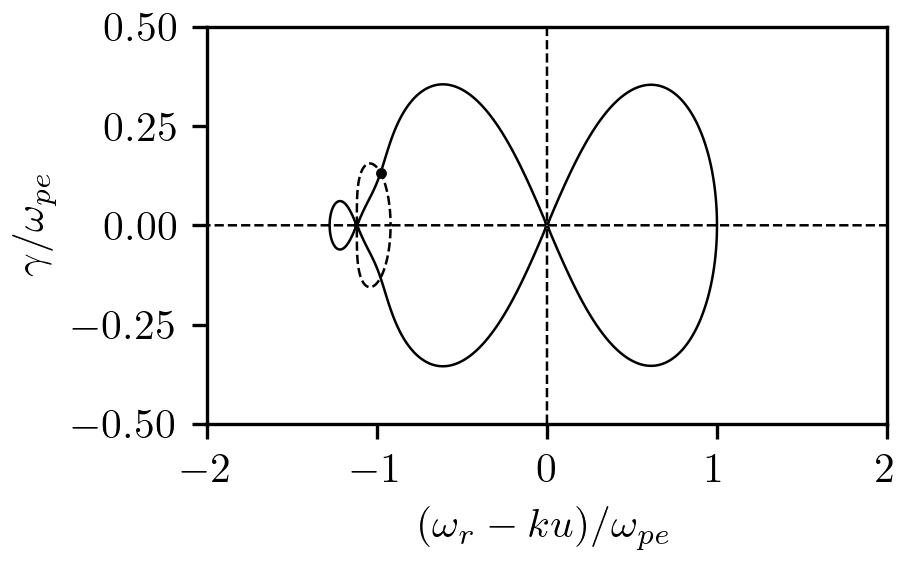}}
    \subfloat[Kinetic, $u/u_0=1$]{\includegraphics{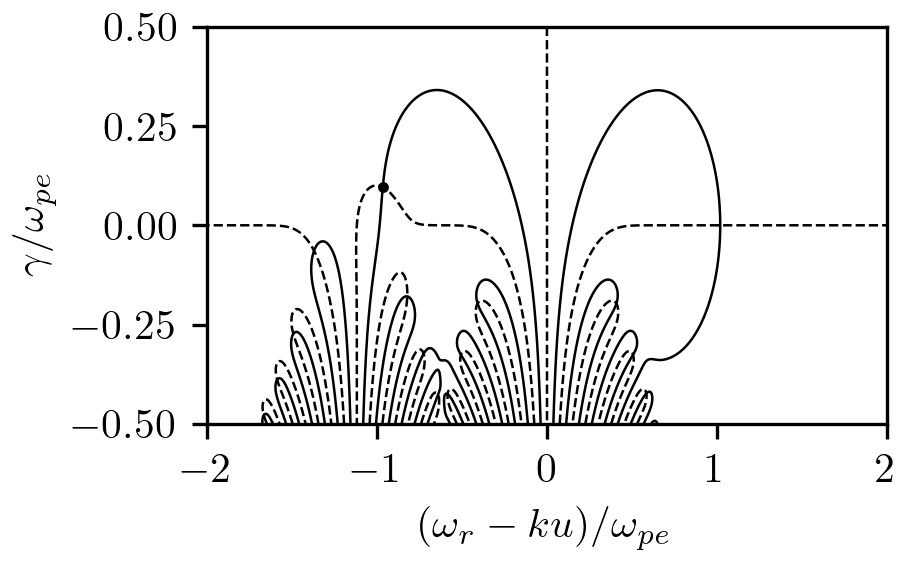}}

    \subfloat[Fluid, $u/u_0=0.75$]{\includegraphics{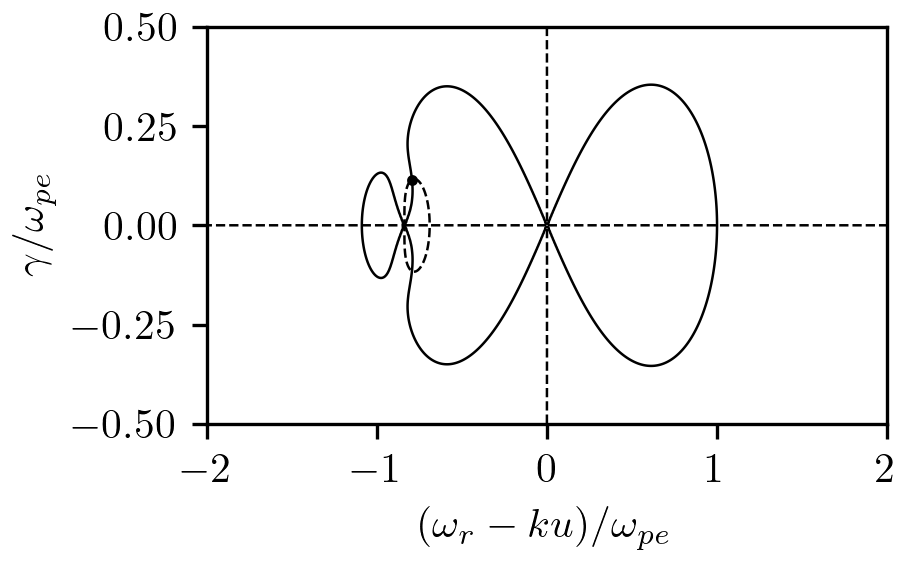}}
    \subfloat[Kinetic, $u/u_0=0.75$]{\includegraphics{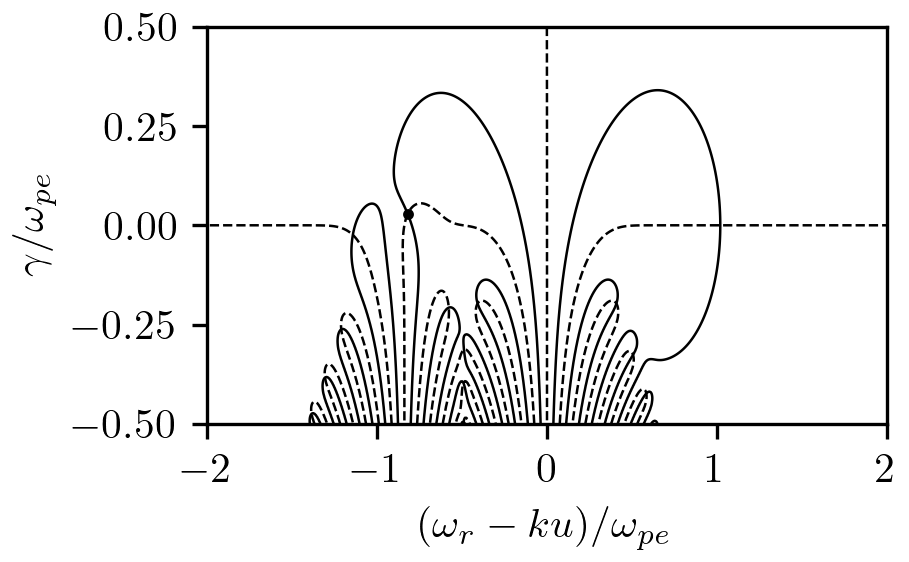}}

    \subfloat[Fluid, $u/u_0=0.5$]{\includegraphics{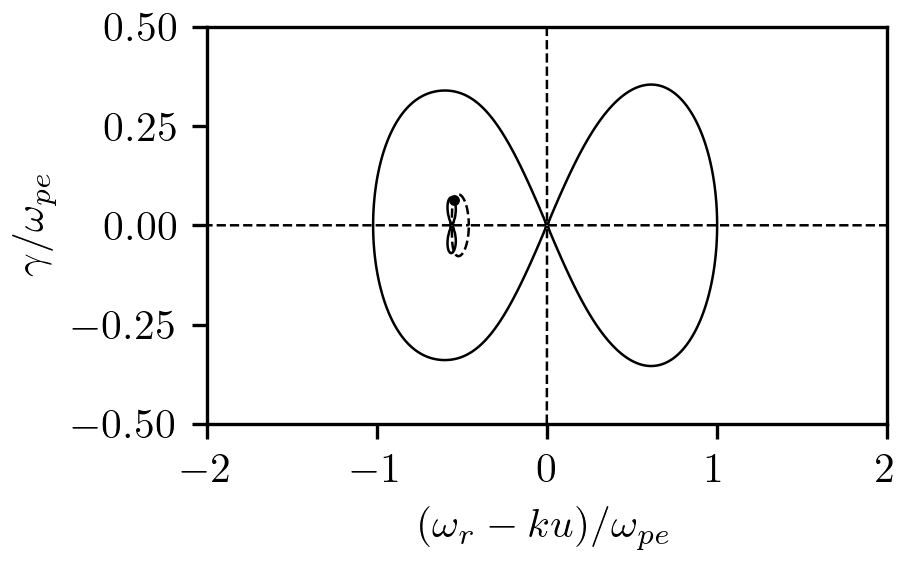}}
    \subfloat[Kinetic, $u/u_0=0.5$]{\includegraphics{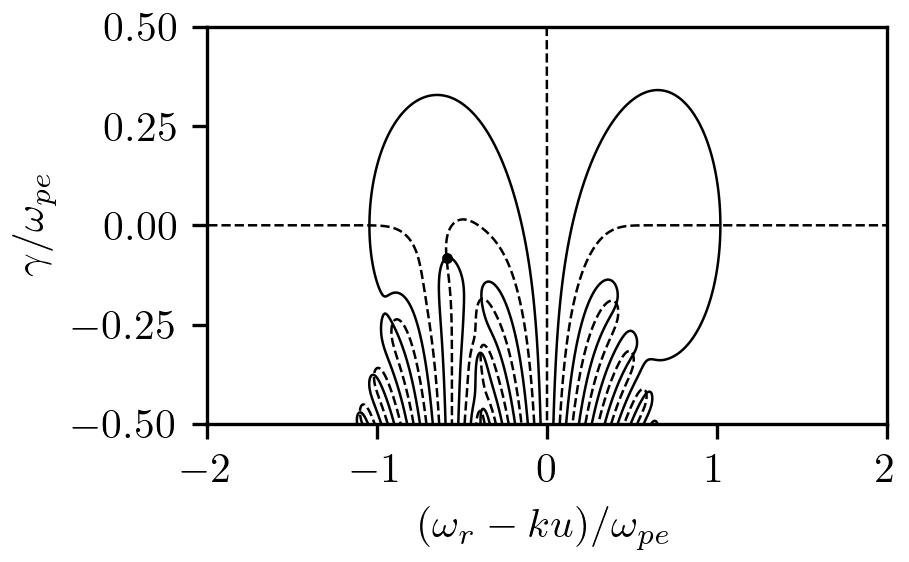}}

    \subfloat[Fluid, $u/u_0=0$]{\includegraphics{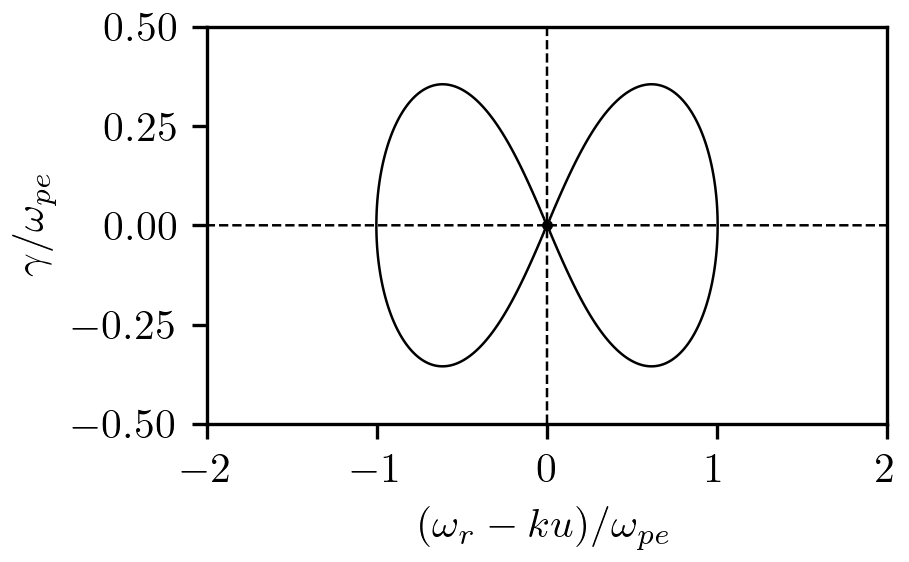}}
    \subfloat[Kinetic, $u/u_0=0$]{\includegraphics{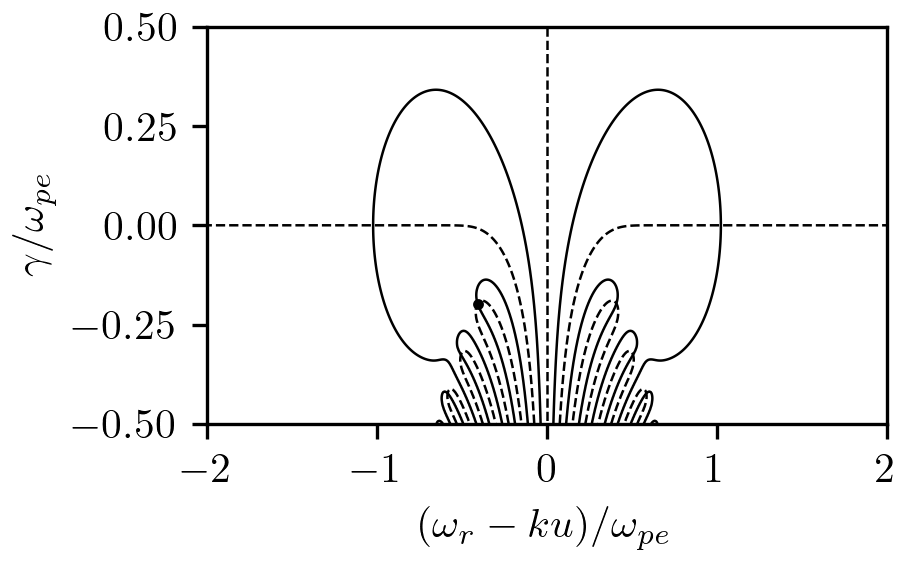}}

    \caption{Contour plots of the dispersion relation in \cref{eq:disprel_asymmetric} (left) and \cref{eq:disprel_asymmetric_kinetic} (right) for the Buneman case for $k=1.12$ (the fastest-growing mode). The solid contours are where $\Re{\varepsilon}=0$, and the dashed contours are where $\Im{\varepsilon}=0$. The roots $\omega_r+i\gamma$ where $\varepsilon(\omega_r+i\gamma, k)=0$ are thus at the intersections between solid and dashed contours. The $x$-axis is shifted by the slope $ku$, i.e., Doppler-shifted to the frame of the electrons, to keep the contours due to the electrons centered regardless of $u$. The black dot indicates the nominal mode, i.e., the mode that grows fastest at $u/u_0=1$.}
    \label{fig:growth}
\end{figure*}

Inspired by the thesis of \citet{cagas}, we use contour plots to identify solutions of the dispersion relations, but for a range of different electron drift velocities $u\in[0,u_0]$. $u_0$ is the inital drift velocity in accordance with \cref{tab:parameters}. Considering first the Buneman case, \cref{fig:growth} shows contour plots for the fluid dispersion relation in \cref{eq:disprel_asymmetric} on the left, and the full kinetic dispersion relation in \cref{eq:disprel_asymmetric_kinetic} on the right. We do not actually account for an oscillating ion drift in the equations, but as shown in \cref{sec:ie_oscillations}, the relative drift oscillates in the same way for electron-only and ion--electron oscillations, with the consequence that the growth rates will be the same regardless. All plots are for the value of $k$ used in the simulations, with real and imaginary parts of $\omega$ on the axes (do not be confused by the shift in the $x$-axes; it prevents the contours from sliding left--right as we change $u$). The solid lines are contours where $\Re{\varepsilon(\omega, k)}=0$, whereas the dashed lines are contours where $\Im{\varepsilon(\omega, k)}=0$. At the solutions, i.e., when $\varepsilon(\omega, k)=0$, the solid and dashed contours intersect (incidentally, for the fluid case they also intersect at the singularities of \cref{eq:disprel_asymmetric}). For the case of $u/u_0=0$, (ordinary) Langmuir oscillations are predicted at $\omega_r/\omega_{pe}=\pm 1$, but with no (visible) growth or damping. For the kinetic theory, the left ``fan'' is due to the plasma dispersion function for the ions, whereas the right ``fan'' is due to the electrons. The kinetic contour plots also show many modes (intersections) in the lower middle part of the plot, but these are heavily Landau damped and will not emerge in the practice. Hence, the fluid theory is a good approximation to the kinetic theory in this case. The upper two subfigures correspond to the initial relative drift velocity of the simulations. The fastest-growing mode (the one with largest imaginary part) is indicated with a dot in \cref{fig:growth}, and although it is slightly lowered due to Landau damping in the kinetic case, the fluid theory is still a reasonable approximation. Naturally, the coordinates of the dot coincide with the values found in \cref{sec:buneman}, and listed in \cref{tab:buneman} for the fluid and kinetic theories. However, as seen in the middle panes, the maximum growth rate diminishes as the relative drift velocity tends towards zero, and for the kinetic dispersion relations, it even becomes negative. This is due to Landau damping when the two distributions overlap. Since the Landau damping is large for significant parts of the period of the current-driven Langmuir oscillations, it is important to use the full kinetic dispersion relation when computing the averaged growth rate.

We tabulate the growth rate $\gamma(u)$ for different drift velocities $u\in[0,u_0]$, with a stepsize of $\Delta u/u_0=0.01$. We start at $u=u_0$, where an initial guess of the fastest growing mode is easily found by visual inspection of the dispersion relation. We then use the secant method (\texttt{scipy.optimize.newton} in SciPy \cite{scipy}) to find a more accurate value. For each lower value of $u$, we use the previous root as an initial guess to the secant method, to track the root as the drift velocity decreases. An animation of the procedure is available in the supplementary material \cite{supplementary}.

Knowing that $u(t)$ oscillates according to \cref{eq:ie_oscillations_relative}, and having tabulated values for $\gamma(u)$, we can now derive an expression for an \emph{equivalent growth rate} $\bar\gamma$.
Let one period $T=2\pi/\omega_0$ of Langmuir oscillations be discretized into time instants $t_k=k\Delta t$, $k=0,...,N$, with $\Delta t$ small enough that $\gamma$ does not change much during a timestep, i.e., $\Delta t\ll T$. During one timestep, e.g., from time $t_{k-1}$ to $t_k$, all linearized quantities will grow approximately by a factor $e^{\gamma(u(t_k))}$.
We may thus define an equivalent growth rate $\bar\gamma$ by chaining together the growth factors over all timesteps $\Delta t$ during an entire period:
\begin{align}
    e^{\bar\gamma T} = \prod\limits_{k=1}^N e^{\gamma(u(t_k))\Delta t}
    \label{eq:growth_product}
\end{align}
Taking the logarithm of both sides,
\begin{align}
    \bar\gamma = \frac{1}{T} \sum\limits_{k=1}^N \gamma(u(t_k))\Delta t
    \label{eq:growth_sum}
\end{align}
and taking the limit $\Delta t\rightarrow 0$ we arrive at the Riemann integral, which shows that the equivalent growth rate is indeed an average:
\begin{align}
    \bar\gamma = \frac{1}{T} \int\limits_0^T \gamma(u(\tau)) \,\dd\tau
    = \frac{4}{T} \int\limits_0^{T/4} \gamma(u(\tau)) \,\dd\tau.
    \label{eq:growth_integral}
\end{align}
The latter equality follow due to symmetry: $\gamma(-u)=\gamma(u)$ due to the geometry of the problem, and for $u(t)$ given by \cref{eq:ie_oscillations_relative}, there is odd symmetry about the time $T/4$.

As mentioned, $\gamma$ is obtained by the secant method for different values $u$ at a grid with regular spacing $\Delta u$ instead of $\Delta t$. It is useful, then, to discretize the integral with the trapezoidal rule \cite{suli}, assuming non-uniform spacing $t_k-t_{k-1}$,
\begin{align}
    \bar\gamma\approx\frac{4}{T}\sum_{k=1}^N\frac{\gamma(u_{k-1})+\gamma(u_k)}{2}(t_k-t_{k-1}),
    \label{eq:growth_trapezoidal}
\end{align}
where $u_k=u_0\cos(\omega_0 t_k)$. Alternatively, since $u(t)$ is strictly monotonic in $[0,T/4]$, we can replace $t_k$ in the above equation with $\arccos(u_k/u_0)/\omega_0$. The advantage of the latter approach, is that we can use a table for $\gamma(u)$ that is uniformly spaced in $u$.

We have seen how the growth rate $\gamma$ of the nominal mode, i.e., the one that grows fastest when $u=u_0$, gradually moves downwards and towards the right in the right-hand plots in \cref{fig:growth}. If we numerically integrate the growth rate of that mode using \cref{eq:growth_trapezoidal}, we actually arrive at a negative effective growth rate $\bar\gamma$ (listed in \cref{tab:buneman}). Clearly, this cannot be the mode that follows the minima in \cref{fig:buneman}, since that must have a positive growth rate. We hypothesize that this growth occurs by ``mode hopping''. When the initial mode decreases to a lower growth rate than some other nearby mode, the evolution of the physical system will near-instantaneously switch to another, faster-growing mode. This means that the unstable mode not only switches growth rate, but also frequency. Although this may seem unlikely at first, we remind the reader that the dynamics due to $\omega_r$ are much slower than that due to $\gamma$, this being a strong instability, as well as due to the current-driven Langmuir oscillations. Given the ``insignificance'' of $\omega_r$ in the dynamics, as well as the complexity involved, we do not find it inconceivable that such ``mode hopping'' may occur. As can be seen in \cref{fig:growth}, whenever the nominal mode has negative growth rate $\gamma$, there is always another mode which has approximately zero growth rate. To account for the mode hopping, we repeat the numerical integration using \cref{eq:growth_trapezoidal}, but for simplicity we simply set $\gamma=0$ whenever $\gamma<0$ for the nominal mode. The result of this computation is also given in \cref{tab:buneman}, and remarkably, it is within $3\%$ of the simulation results.

\begin{figure}
    \subfloat[Kinetic, $u/u_0=1$]{\includegraphics{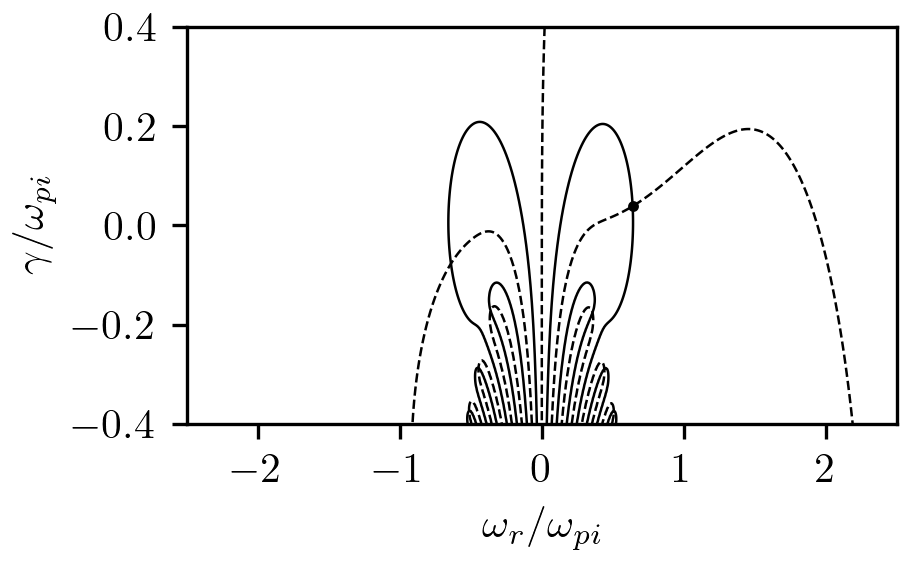}}

    \subfloat[Kinetic, $u/u_0=0.5$]{\includegraphics{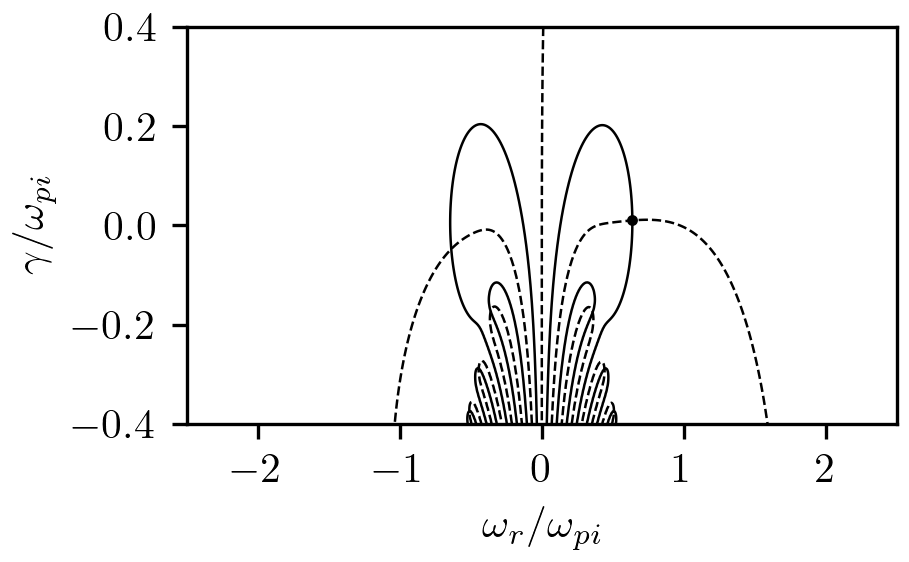}}

    \subfloat[Kinetic, $u/u_0=0$]{\includegraphics{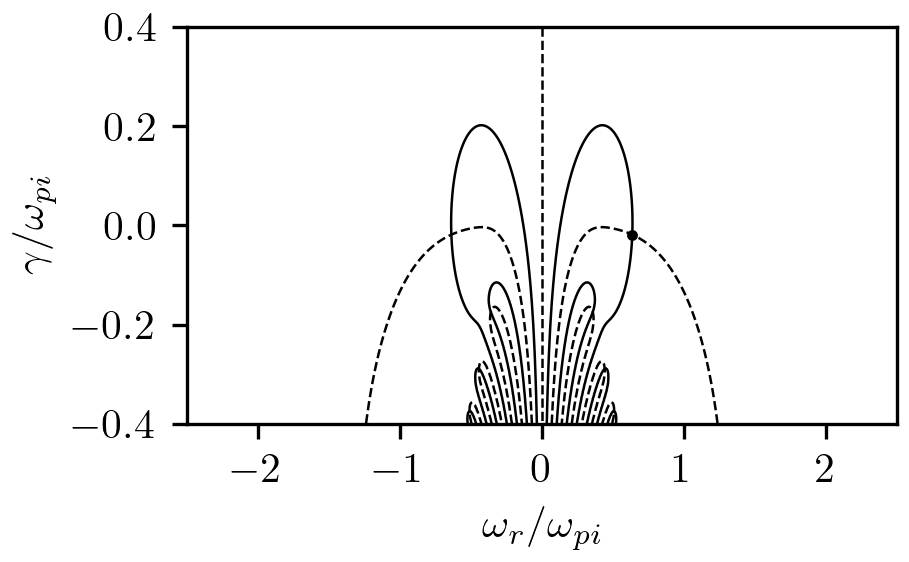}}

    \caption{Contour plots of the dispersion relation in \cref{eq:disprel_asymmetric_kinetic} for the ion acoustic case for $k=1.94$ (the fastest-growing mode). The solid contours are where $\Re{\varepsilon}=0$, and the dashed contours are where $\Im{\varepsilon}=0$. The roots $\omega_r+i\gamma$ where $\varepsilon(\omega_r+i\gamma, k)=0$ are thus at the intersections between solid and dashed contours. The black dot indicates the nominal mode, i.e., the mode that grows fastest at $u/u_0=1$.}
\label{fig:growth_ionac}
\end{figure}

\subsection{The ion acoustic case}
Repeating the procedure for the ion acoustic case, we start with contour plots of \cref{eq:disprel_asymmetric_kinetic}, for the parameters of the ion acoustic case in \cref{tab:parameters}, and the wavenumber given in \cref{tab:ionac}. Again, we do this for steps of $\Delta u/u_0=0.01$, and an animation is available in the supplementary material \cite{supplementary}. Three of the frames are included in \cref{fig:growth_ionac}. What we observe is mostly the ``fan''-shaped plasma dispersion function due to the ions, whereas the electron plasma dispersion function is on a much larger scale in the plots due to the larger thermal speed. Its presence manifests in the way it ``modifies'' the roots of the ion plasma dispersion function, however.

The two intersections furthest up in the plots correspond to the positive and negative branch in \cref{fig:disprel_ionac}. $\omega_r$ of these two modes do not depend much on the relative velocity $u$, in accordance with \cref{eq:ionac_real_roots}. Because of this, the discrepancy in $\omega_r/\omega_{pi}$ between theory and simulations in \cref{tab:ionac} can hardly be explained by drift-averaging. The result of calculating the averaged growth $\bar\gamma$ for the ion acoustic case using \cref{eq:growth_trapezoidal} is given in \cref{tab:ionac}. In this case there is no nearby modes of higher growth rate anytime during the change of $u$, so no mode hopping is assumed. Unfortunately, the drift-averaged growth rate do not agree well with simulation results.
 \section{Conclusion}
\label{sec:conclusion}

We have shown that the configurations normally believed to lead to the Buneman or ion acoustic instability, i.e., the ones depicted in \cref{fig:config_asymmetric,fig:config_ionac}, instead lead to strong oscillations at the plasma frequency in the drift velocity. The same is true for the electron--electron two stream instability, when asymmetric in the velocity space. These oscillations have not only been seen in accurate numerical solutions of the Vlasov-Maxwell equations, but also been predicted by theory, by us and by others. It is of little doubt that these oscillations take place, and that they severely affect the evolution these oscillations have, compared to if the drift velocity did \emph{not} oscillate.

We \emph{hypothesize} that at least two-stream instabilities (including the Buneman case) can co-exist with the current-driven Langmuir oscillations, though possibly at a different growth rate. This because there appear to be a linear slope in the minima of \cref{fig:energy}. For the two-stream case with current, the slope in minima is exactly as predicted by textbooks, whereas for the Buneman case, the growth rate is diminished. This can be qualitatively understood from the fact that the Landau damping increases when the overlap between the electron and ion distributions increases. We have also proposed a method of averaging the growth rate over different drift velocities to account for this varying degree of Landau damping, to good agreement with the simulation results. It is also clear that the oscillation changes the behaviour of the would-be ion acoustic instability. We observe oscillations in the minima which have a frequency in the ballpark of ion acoustic instabilities, but we are not able to account for the damping that occur, instead of growth, with the drift averaging method.

We refrain from concluding much about the domain of validity for the drift-averaging method at this point. It seems a reasonable method for capturing the effects due to varying Landau damping, and seemed to work well in the Buneman case. Nonetheless, there are several effects it does not account properly for, since it is based on a dispersion relation which does not properly account for the oscillations in the unperturbed state. Particularly, the traditional disperion relation assumes perturbations $\delta u\ll u(t)$, which is not the case during zero-crossings of $u(t)$, and a zero unperturbed electric field. Maybe this is why it did not work as well for the ion acoustic case. In the same way we built on past work, there is also plenty of possibilities to improve on our work, and attempting a dispersion relation successfully accounting for the Langmuir oscillations from the start would definitiely give a more complete picture. Another possible improvement would be to describe how current-driven Langmuir oscillations may be Landau dampened when the uniformity is non-perfect, for instance when they co-exist with traditional Landau oscillations.

Finally, and most importantly, our work is all in vain if it does not get read and used for the greater good of humanity, though presumably only small steps at a time. These are controversial final words in a scientific paper, but nonetheless true. We hope to inspire our readers to make conscious decisions in how they apply their skills, including the knowledge gained from our work.

 \section*{Acknowledgment}
This work received funding from the European Research Council (ERC) under the European Union's Horizon 2020 research and innovation program (Grant Agreement No. 866357, POLAR-4DSpace). It also received funding from the Research Council of Norway (RCN), grant number 275653.

S.M. gratefully acknowledge a couple of discussions with Hans Pécseli about the works of \citet{sauer2015} and the Buneman instability. In particular, Hans Pécseli insisted that the Poisson equation should be adequate for electrostatic phenomena, which eventually led to \cref{sec:helmholtz}.
 \section*{Contribution statement}
Sigvald Marholm rediscovered the current-driven Langmuir oscillations previously discussed by \citet{baumgaertel,sauer2015,sauer2016} when trying to simulate streaming instabilities, and used it to explain the observed behavior. He also discovered that instabilities can be traced in the minima of the energy in \cref{fig:energy}, and proposed the drift-averaging method. S.~M. has derived the equations, written the text, and carried out the simulations.

Sayan Adhikari has been learning about instabilities and the Gkeyll code together with S.~M. He has taken part in discussions on a near-daily basis, and continually provided feedback on the text and the simulations, with critical questions that helped shape the project. He also made the first version of \cref{fig:disprel_twostream} using a polynomial root finder on \cref{eq:disprel_asymmetric}.

Wojciech J. Miloch originally suggested simulating the ion acoustic instability as a stepping stone towards understanding ionospheric processes. When that lead to then-unexpected oscillations, he suggested trying the Buneman instability instead.
 \section*{Data availability statement}
Input files to all Gkeyll simulations can be found in the supplementary material \cite{supplementary}. The simulation results can be reconstructed by running the simulations, which should take no more than a few hours. The scripts used for post-processing, and finding the roots of dispersion relations, are throw-away code. The steps taken to obtain these results should be described in sufficient detail in the paper to be repeated.

\appendix
\section{Growth rate of two-stream instabilities}
\label{app:growthrate}
Finding the growth rate of the electron--electron two-stream instability is easiest in the symmetric case, when substituting $\omega\rightarrow\omega+0.5ku$ in \cref{eq:disprel_asymmetric}. Multiply through by the denominator and rearrange:
\begin{align}
    (\bar\omega^2-\bar k^2)^2 - 2(\bar\omega^2-\bar k^2) - 4\bar k^2 = 0
\end{align}
where $\bar\omega = \omega/\omega_{pe}$ and $\bar k=0.5ku/\omega_{pe}$. This is a second order equation in $\bar\omega^2-\bar k^2$, with the solution
\begin{align}
    \bar\omega^2=\bar k^2+1\pm\sqrt{1+4\bar k^2}.
\end{align}
Since $\bar\omega^2\in\mathbb R$, $\bar\omega$ is either purely real or imaginary. Assuming it to be imaginary, and returning to the asymmetric frame of reference results in the \cref{eq:fastest_growing_twostream}.

For the Buneman instability the two frequencies are different. Following the approach in \cite{treumann_advanced}, we define
\begin{align}
    \omega_+ = ku+\omega_{pe}, \\
    \omega_- = ku-\omega_{pe},
\end{align}
(corresponding to the fast and slow branches, respectively) such that 
\begin{align}
    (\omega-\omega_-)(\omega-\omega_+) = (\omega-ku)^2-\omega_{pe}^2.
\end{align}
Multiplying \cref{eq:disprel_asymmetric} throughout by $\omega^2(\omega-ku)^2$ and using the above identity we can write the dispersion relation as
\begin{align}
    (\omega-\omega_-)\omega^2=\frac{\omega_i^2(\omega-ku)^2}{\omega-\omega_+}.
    \label{eq:disprel_pm}
\end{align}
Since the fastest-growing mode occur approximately at the intersection between the slow branch and positive ion brach (see \cref{fig:disprel}), $ku-\omega_{pe}\approx\omega_i$. Moreover, since $\omega_i\ll\omega_{pe}$, $ku\approx\omega_{pe}$, which is the same as in \cref{eq:fastest_growing_buneman}, and
\begin{align}
    \omega_+ \approx 2\omega_{pe} ,\quad
    \omega_- \approx 0.
\end{align}
Substituting this into \cref{eq:disprel_pm}, and using $\omega\ll\omega_{pe}$ (since it is in the order of $\omega_i$), leads to
\begin{align}
    \omega^3 \approx -\frac{1}{2}\frac{m_e}{m_i}\omega_{pe}^3.
\end{align}
Perhaps the easiest way to solve this is to write both sides as polar form complex numbers, and equating the magnitude and argument of both sides. The solutions are
\begin{align}
    \frac{\omega}{\omega_{pe}} = 
    \left(\frac{1}{2}\frac{m_e}{m_i}\right)^\frac{1}{3}
    \exp\left(i\frac{\pi}{3}(1+2n)\right), \quad n=0,1,2.
\end{align}
Only for $n=0$ is there growth, and that can also be written like in \cref{eq:fastest_growing_buneman}.
 \section{Plasma dispersion function}
The plasma dispersion function is defined as \cite{fitzpatrick,cagas}
\begin{align}
    Z(\zeta) &= \frac{1}{\sqrt{\pi}}
    \int\limits_{-\infty}^\infty \frac{e^{-t^2}}{t-\zeta} \,\dd t \nonumber \\
    &= i\sqrt{\pi} e^{-\zeta^2}(1+\erf(i\zeta)).
\end{align}
The latter form is convenient when implementing the plasma dispersion function in scripts.
Its derivatives can be obtained through the following recurrence relations
\begin{align}
    Z'(\zeta) &= -2(1+\zeta Z(\zeta)) \nonumber\\
    Z^{(n)}(\zeta) &= -2(Z^{(n-2)} + \zeta Z^{(n-1)} Z(\zeta))
\end{align}
 
\end{document}